\documentclass[prx, twocolumn, secnumarabic,nobalancelastpage,amsmath,amssymb,nofootinbib,floatfix]{revtex4-2}

\usepackage{comment}
\usepackage{graphicx}
\usepackage{bm}
\usepackage{bbm}
\usepackage{bbold}
\usepackage{amsmath}
\usepackage{braket}
\usepackage{color}
\usepackage[normalem]{ulem}
\usepackage{soul}
\usepackage{xcolor}
\usepackage{blindtext}
\usepackage{lipsum}

\setstcolor{red}

\global\long\def\bra#1{\left\langle #1\right|}
\global\long\def\ket#1{\left|#1\right\rangle }
\global\long\def\braket#1#2{\left\langle #1|#2\right\rangle }
\global\long\def\ketbra#1#2{\left|#1\right\rangle \left\langle #2\right|}

\global\long\def\dag{^{\dagger}}

\global\long\def\hamil{H}
\global\long\def\tens{\otimes}

\global\long\def\a{\mathcal{\alpha}}

\global\long\def\a{\mathcal{\alpha}}

\begin{document}

\title{Conditional not displacement: fast multi-oscillator control with a single qubit}

\author{Asaf A. Diringer}
\author{Eliya Blumenthal$^1$}
\author{Avishay Grinberg$^1$}
\author{Liang Jiang$^2$}
\author{Shay Hacohen-Gourgy$^1$} 
\affiliation{$^1$Department of Physics, Technion - Israel Institute of Technology, Haifa 32000, Israel}
\affiliation{$^2$Pritzker School of Molecular Engineering, University of Chicago, Chicago, Illinois 60637, USA}
\begin{abstract}
Bosonic encoding is an approach for quantum information processing, promising lower hardware overhead by encoding in the many levels of a harmonic oscillator mode. Scaling to multiple modes requires weak interaction for independent control, yet strong interaction for fast control. Applying fast and efficient universal control on multiple modes remains an open problem. Surprisingly, we find that displacements conditioned on the state of a \textit{single} qubit ancilla coupled to multiple harmonic oscillators are sufficient for universal control. We present the conditional-no operation concept, which can be used for reducing the duration of entangling gates. Within this guiding concept, we develop the conditional not displacement control method which enables fast generation and control of bosonic states in multi-mode systems weakly coupled to a single ancilla qubit. Our method is fast despite the weak ancilla coupling. The weak coupling in turn allows for excellent separability and thus independent control. We demonstrate our control on a superconducting transmon qubit weakly coupled to a multi-mode superconducting cavity. We create both entangled and separable cat-states in different modes of the multi-mode cavity, showing entangling operations at low cross-talk while maintaining independent control of the different modes. We show that the operation time is not limited by the inverse of the coupling rate, which is the typical timescale, and we exceed it by almost 2 orders of magnitude. We verify our results with an efficient method for measurement of the multi-mode characteristic function which employs our conditional not displacement. Our results inspire a new approach toward general entangling operations and allow for fast and efficient multi-mode bosonic encoding and measurement.
\end{abstract}

\maketitle

Bosonic codes are an approach in which quantum information is redundantly encoded in the many levels of a quantum harmonic oscillator. Unlike the conventional approach to quantum error correction, which relays on multiple two-level systems, bosonic code are hardware efficient as they allow for errors to be corrected already at the level of a single quantum element~\cite{gottesman_encoding_2001,chuang_bosonic_1997,lund2008fault}. Concatenating these bosonic qubits into a larger error-correcting code may provide a path to robust quantum computation with reduced hardware overhead~\cite{guillaud2019repetition,menicucci2014fault,chamberland2022building}. Significant progress has been made in two leading platforms, circuit QED and Trapped Ions, where small error correction protocols for stabilizing a bosonic qubit have been demonstrated.~\cite{ofek_extending_2016,campagne2020quantum,de2022error}. Photonic systems are also finding new and improved schemes for generating such states~\cite{dahan2022creation, hastrup2020deterministic,lewenstein2021generation}.

To achieve control over a single harmonic oscillator it must be coupled to an ancilla. Universal control can be obtained by a combination of unconditional and conditional operations~\cite{krastanov_universal_2015,PhysRevLett.115.137002,fosel2020efficient,kudra2022robust, eickbusch2022fast}, or numerical pulse optimization~\cite{khaneja2005optimal,heeres_implementing_2017}. Extending to multiple oscillators, universal control means any state in the joint Hilbert space of the oscillators can be prepared and any unitary can be implemented. Hence, one must find ways to control each harmonic mode and also couple them. How to achieve universal control of multiple harmonic oscillators in a fast and efficient manner while minimizing hardware overhead and cross-talk remains a gap to be filled.

Typically, separate ancillary systems are used for entangling operations between different modes, and additional ancillas for control and measurement of the different modes. This complicates matters by introducing additional sources of decoherence and increasing the hardware overhead. While coupling a single ancilla to all modes would be simpler, conditional operations with a single ancilla do not trivially extend to multiple modes as frequency crowding becomes prohibitive~\cite{naik2017random,chakram2022multimode}. Furthermore, conditional operations are slow due to the need for frequency selectivity, which limits their speed to the inverse of the interaction rate. While faster operations are preferred to reduce decoherence effects, increasing interaction strength to speed up the process is not ideal as it causes unwanted interactions. Therefore, strong ancilla interactions are beneficial for faster entangling operations, but for independent and cross talk suppressed control, a weak ancilla interaction is preferable.
 
\begin{figure*}[t!]
    \includegraphics[width=\linewidth]{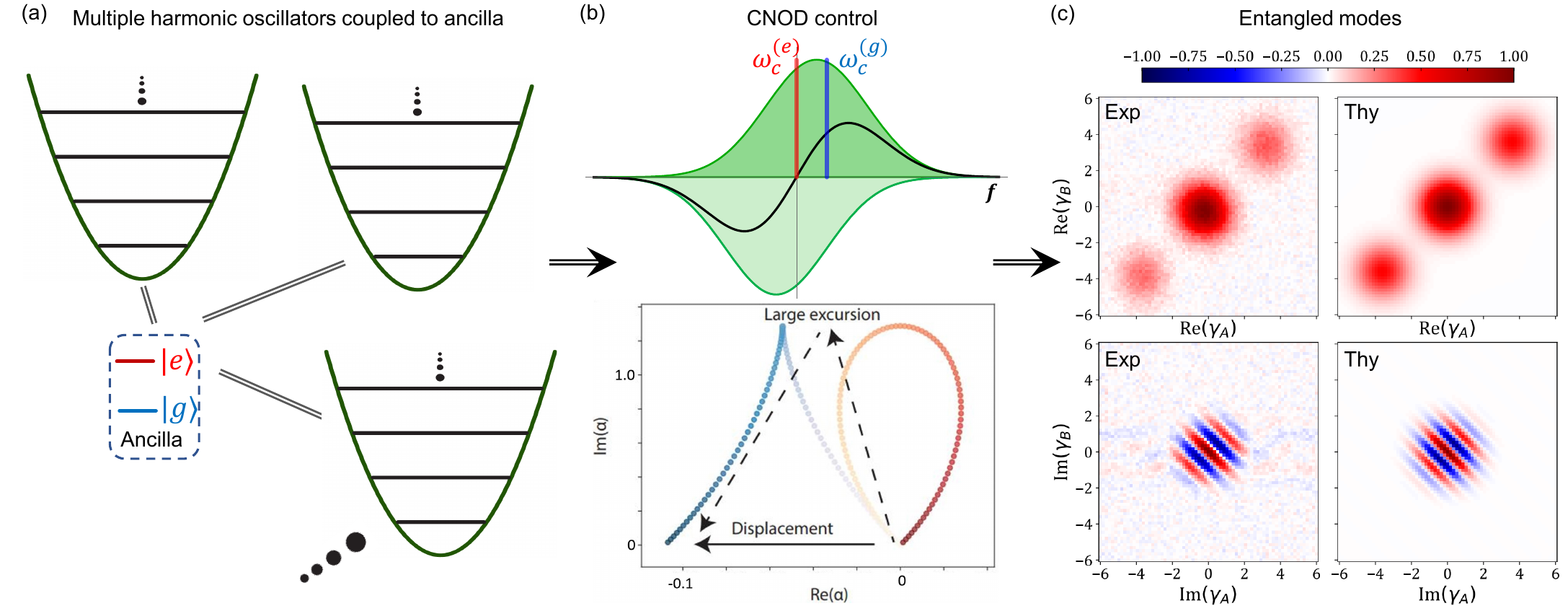}
\caption{\textbf{Conditional NOt Displacement (CNOD) multi-mode control} (a) Illustration of a single qubit ancilla coupled to multiple Bosonic modes. (b) \textit{Top} - Schematic illustration of the Fourier decomposition of the anti-symmetric pulse, these pulses are used to compose the CNOD. The anti-symmetric pulse is a sum of two Gaussians (green), which have the same amplitude but opposite phases, and their center frequencies are equally detuned from either side of $\omega_{c}^{(e)}\equiv \omega_{c}-\chi$ (frequency of the EM mode when the ancilla is excited). The Fourier transform of the sum of the two Gaussians (qualitatively represented by the black line) is anti-symmetric with respect to  $\omega_{c}^{(e)}$, resulting with a node at that frequency. The result is a no-displacement conditioned on the excited state of the ancilla. \textit{Bottom} - Two simulated trajectories of the EM mode phase space show the dynamics of the pulse. The initial state is cavity vacuum, where the ancilla is in the exited- (red) and ground- (blue) state with the time evolution portrayed by the color gradient. The dynamics show that initially, as the anti-symmetric pulse is applied, the oscillator is largely displaced independently of the state of the ancilla. The node present at the oscillator frequency corresponding to the excited state of the ancilla leads to a fully destructive interference. When the ancilla is at the ground state the interference is not fully destructive and a finite displacement occurs. (c) Joint characteristic tomography of Bell-cats (left) compared with ideal state (right). 2D cuts of the 4D joint characteristic functions $[\mathcal{C}_{J} (\gamma_{A},\gamma_B )]$ of the Bell-cats $\ket{\psi_+}$. The cuts are along the planes, Re$(\gamma_A)$-Re$(\gamma_B)$ and Im$(\gamma_A)$-Im$(\gamma_B)$, for $\alpha_A\simeq\alpha_B \simeq 1.7$).}
    \label{fig:AS pulse}
\end{figure*}

We show the counter-intuitive result that we can perform multi-mode universal control through displacements conditioned on the state of a \textit{single} ancilla qubit coupled to multiple harmonic oscillators, as illustrated in Fig.~\ref{fig:AS pulse}a.

We use our conditional-no concept to construct the Conditional NOt Displacement (CNOD) gate for an oscillator dispersively coupled to a qubit. The conditional-no concept is general and can be understood as follows. Rather than applying a narrow-band pulse at a selected frequency for realizing a controlled operation, nodes are introduced to select where the operation does not occur. The \emph{no} operation is achieved by designing a zero amplitude at the selected frequencies in the Fourier spectrum of the control pulse. This approach alleviates the requirement for a narrow pulse bandwidth and is extendable to multiple frequencies. Consequently, the CNOD duration is not limited by the typical time scale $1/\chi$, where $\chi$ is the cross-Kerr coupling strength associated with the frequency shift of the oscillator dependent on the ancilla state. 


While such speedup was recently demonstrated for a single oscillator through the Echoed Conditional Displacement (ECD) gate~\cite{eickbusch2022fast}, the CNOD displays different dynamics, as detailed in the supplementary, and is derived from our more general concept of conditional-no operation. Being a general concept it can be transferred to other control systems by replacing frequency selectivity by frequency anti-selectivity, which may alleviate control constraints.
 
 

Insensitive to both frequency crowding and to the different values of the dispersive coupling strength, the  CNOD naturally extends to the multi-mode setup. So far, entanglement between modes has been demonstrated in the relatively large coupling regime, at a slow rate and with a few ancillary systems~\cite{wang_schrodinger_2016,gao2019entanglement,ma2020manipulating} or in the weak coupling regime using a single ancilla, but at a slow rate using the Zeno effect~\cite{chakram2022multimode}. 
By combining the single-ancilla, multiple-oscillator approach with the CNOD gate we obtain an efficient method for multi-mode universal control which can avoid cross-talk by operating at the weak coupling regime. Furthermore, in systems where displacements are the only conditional operation available, our single ancilla approach enables control of multiple modes~\cite{baranes2023free}.

In our proof-of-concept experiment, we use an ancilla superconducting qubit weakly coupled to a multi-mode superconducting microwave cavity. We demonstrate that the CNOD operation speed is not limited by $1/\chi$, and we exceed it in practice by almost 2 orders of magnitude. Through the generation of both entangled and separable cat-states, we demonstrate multi-mode control at low cross-talk. We also show how the CNOD can be employed to independently map both local and non-local properties onto a single ancilla qubit, which is advantageous for multi-mode tomography.

In typical tomography schemes, the expectation values of non-local operators are obtained by measurements of separate ancillas for each mode. The results are compared for correlations~\cite{wang_schrodinger_2016,gao2019entanglement,ma2020manipulating}, hence multiple ancilla and high-fidelity single-shot ancilla measurements are required. The multi-mode CNOD simplifies tomography. Using the CNOD we map non-local properties onto the ancilla. Since correlations in the direct measurement of the non-local operators are preserved during averaging, both requirements are removed in our method.
We can now construct the specific conditional displacement on a single mode for our circuit QED system. We consider a two-level system of frequency $\omega_q$ dispersively coupled to an EM mode of frequency $\omega_c$. Working in the frame rotating with the ground state of the ancilla and the EM mode, such a system is described by a dispersive Hamiltonian:
\begin{equation}
\hamil_s/\hbar=
-\chi\ketbra{e}{e}a^{\dag}a,\label{eq:stat_hamil}
\end{equation}
where $\chi$ is the coupling strength, $\ket{e}$ is the ancilla excited state, and $a$ is the EM mode lowering operator. We start with a pulse whose Fourier amplitude at the frequency of the EM mode when the ancilla is excited ($\omega_c^{(e)}$) is zero. While any such drive would suffice, we implement the CNOD with anti-symmetric pulses, as shown in Fig.\ref{fig:AS pulse}b. We chose the anti-symmetric shape as it should provide robustness to imperfections in pulse delivery, for example, an amplitude-dependent transfer matrix. These pulses will induce a displacement of the memory mode only if the ancilla is in the ground state, and not in the excited state.
However, as the pulse is applied, an additional conditional rotation of the phase space of the EM mode is induced by the static Hamiltonian. Therefore, the overall system evolution during such a drive is given by the unitary
\begin{equation}
\mathrm{AS}(\alpha,\tau)\equiv\ketbra{g}{g}\mathcal{D}(\alpha/2)+\ketbra{e}{e} \exp(i\chi \tau a^{\dag} a),     
\end{equation}
where $\tau$ is the duration of the anti-symmetric pulse, $ \mathcal{D} (\beta)= e^{\beta a^{\dagger} - \beta^* a}$ is the displacement operator and $\alpha/2$ is the amplitude of the conditioned on-ground displacement induced by the anti-symmetric pulse. 
The conditioned phase space rotation angle $-\chi \tau$ should be very small due to the pulse being much faster than $1/\chi$. However, the operation becomes highly conditional when $\alpha$ is set to a large value. 
This conditional-no operation, as discussed in the supplementary, has various applications.  For bosonic codes, the natural development is an operation equivalent to the standard conditional displacement, thus we want to cancel the conditional rotation.

\begin{figure}[htp!]
    \centering
    \includegraphics[width=\linewidth]{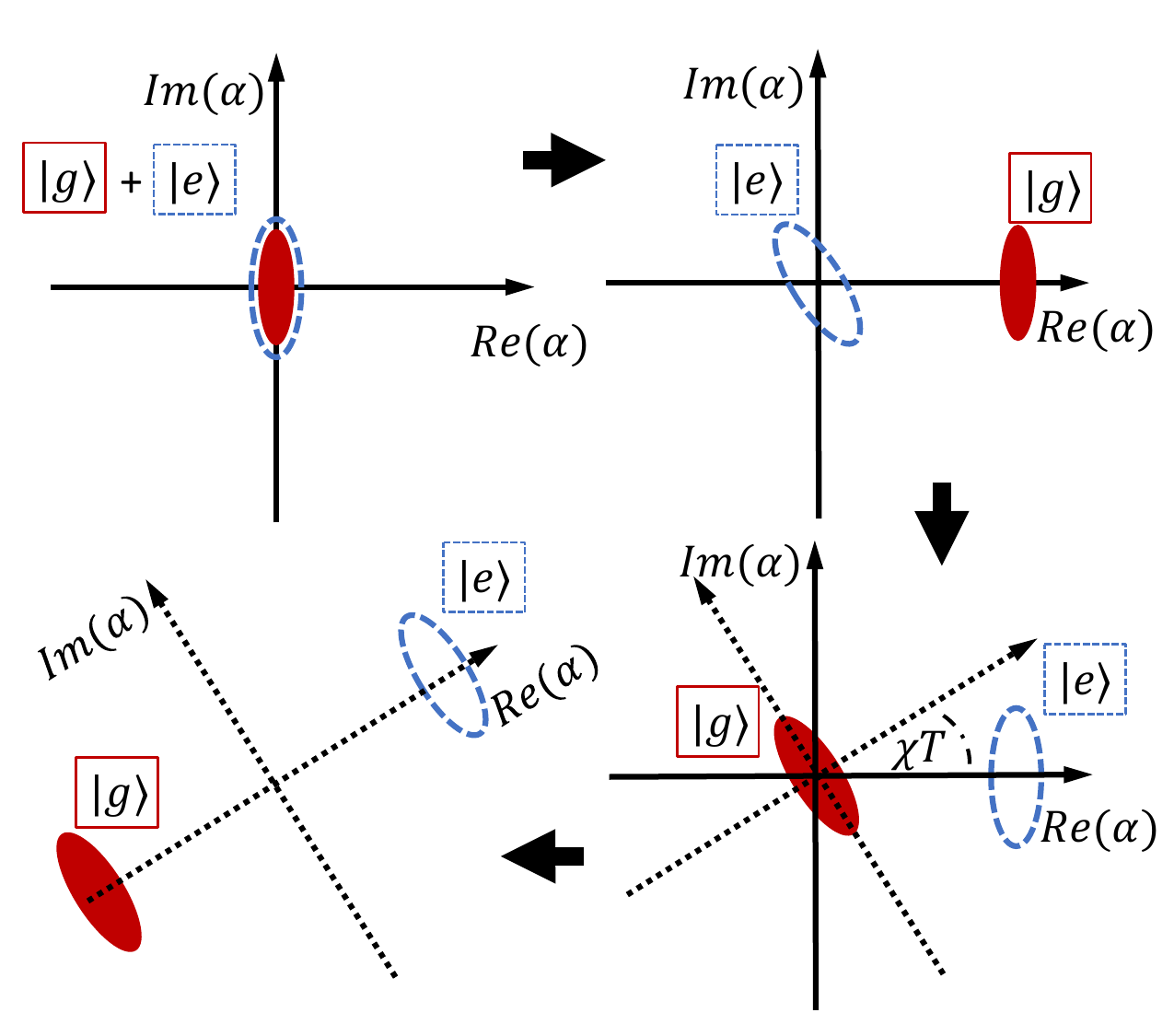}
    \caption{\textbf{CNOD scheme.} 
    Steps in phase space evolution during the $\mathrm{CNOD}$ scheme, shown in the frame rotating at $\omega_c ^{g}$, as described in the main text. We start with an initial state (top left) as a squeezed vacuum to make phase space rotations for an arbitrary initial state visible through the change in orientation of the squeezing axis. The oscillator state corresponding to the excited ancilla state $\ket{e}$ is represented by the full red ellipse while the one corresponding to the ancilla ground state $\ket{g}$ is represented by the dashed blue ellipse. Second, third, and final steps are after the first anti-symmetric pulse (top right), unconditional $\pi$-pulse (bottom right), and a second anti-symmetric pulse followed by a digital axis rotation (bottom left) respectively.}
  \label{fig:CNOD scheme}
\end{figure}

To cancel the state-dependent rotation of the EM mode we apply an unconditional $\pi$ rotation on the qubit ancilla and perform a digital frame rotation of $-\chi\tau$ to the EM mode. The two operations are then followed by a second anti-symmetric pulse, identical to the first up to a minus sign. 
All together, up to a local Z rotation, this procedure results in a displacement of the cavity mode conditioned on the ancilla state. 
\begin{equation}
\mathrm{CNOD}(\alpha)\equiv\ketbra{e}{g}\mathcal{D}(\alpha/2)-\ketbra{g}{e}\mathcal{D}(-\alpha/2)  
\end{equation}
\begin{figure}[htp!]
    \centering
    \includegraphics[width=\linewidth]{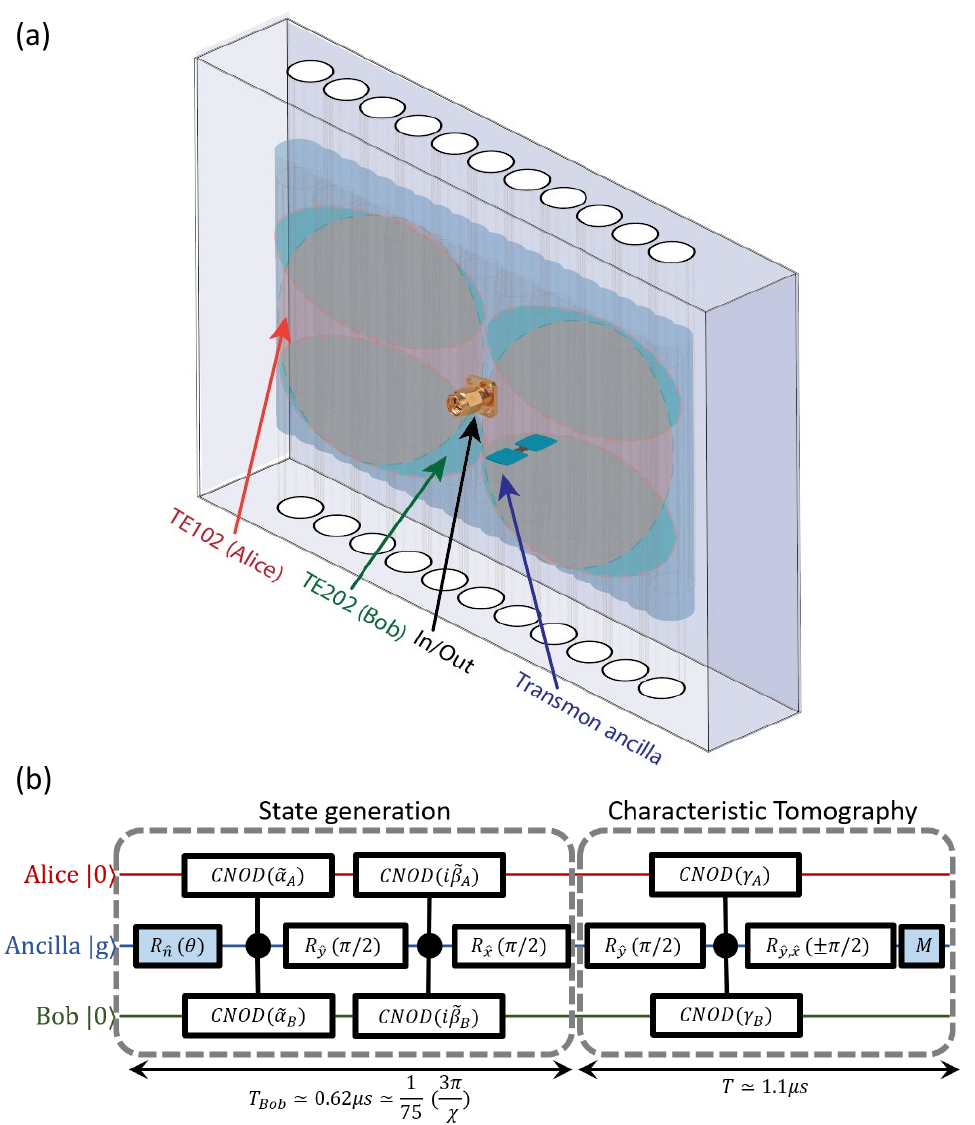}
    \caption{\textbf{Device illustration and protocol for generating cat-states} (a) Schematic of the 3D aluminum flute cavity with a single transmon qubit ancilla inside. The cavity is made of a single slab of high-purity aluminum. Holes (white circles) are drilled from opposite sides, creating a seamless cavity in the center of the aluminum block (blue shaded columns). The cavity has a single pin-antenna (golden connector), which serves as a port for both input and output. The pin-antenna is located at the nodes of the memory modes TE102-Alice (red) and TE202-Bob (green), and at the anti-node of the readout mode TE101 (not shown in illustration). The pin-antenna is used for control of all elements as well as for readout. (b) Microwave control sequence for generating the different cat-states and tomographically reconstructing the characteristic functions, with parameters depending on the target state. The first ancilla rotation $R_{\hat{n}}(\theta)$ determines phase of the cat-state. The first CNODs generate cats entangled to the ancilla, and the following CNODs with $\tilde{\beta_i}$'s set to satisfy $\sum_{i=1,2} \Tilde{\alpha_i}\Tilde{\beta_i}=\pi/2$ disentangle the ancilla from the EM modes. In the final rotation in the tomography: 1) the choice of $\hat{y}$ or $\hat{x}$ determines if we measure the real or imaginary part of the characteristic function respectively, 2) the $\pm$ are interleaved to remove biasing error in the ancilla measurement. Generation times for the Alice-, Bob-, and Bell-cats were 476ns, 620ns, and 1312ns respectively. These include the qubit rotations $\pi$-pulses and $\pi /2$-pulses which were 22ns and 16ns respectively. These generation times give a speed up of 14, 75, and 36 respectively, as compared with $3 \pi / \chi$. $3 \pi / \chi$ is the typical time in the standard method and also does not include the qubit rotation times~\cite{wang_schrodinger_2016}.
    }
    \label{fig:gates and Setup}
\end{figure}

The magnitude $|\alpha|$ of the conditional displacement induced by a CNOD gate depends on the specific characteristics of the anti-symmetric pulse. Regardless of the exact temporal shape, the anti-symmetric pulses ensure that the linear dependence on $\tau$ will be canceled by the first and second halves of the evolution, which leaves the leading contribution to the next quadratic order $\tau^2$. The displacement also scales linearly with driving amplitude, thus it is proportional to  $A \chi \tau^2$, where $A$ is a scaling factor for the amplitude of the drive. For example, to half the pulse duration a factor 4 increase in amplitude is required, which can impose a technical power limitation on the gate speed. It is also important to note that as the accumulated power of the drive increases, the intermediate photon number also increases, which may further constrain the gate speed.
The CNOD method naturally generalizes to a setting with a single ancilla coupled to multiple modes, by applying multiple resonant pulses (each resonant to the corresponding cavity mode). The operation, which is a displacement conditioned on the ancilla, exclusively influences the state of the mode (or modes) to which it is applied. Importantly, it operates independently and is not reliant on the states of any other modes. In a dispersive Hamiltonian, the ancilla does affect other modes, however the effect is echoed by the $\pi$-pulse in the CNOD. As a result, CNOD based multimode control does not require fine tuning of the differnt $\chi$ values or compensating by involving additional modes of the transmon~\cite{wang_schrodinger_2016,chakram2022multimode,stein2023multi}. Furthermore, as the CNOD operates in the small $\chi$ regime the cross-Kerr between modes is negligible, and the CNOD is achieved by applying the anti-symmetric pulse to more modes.

Our experimental setup uses a circuit QED architecture. A transmon type superconducting ancilla is introduced into to a 3D flute style superconducting cavity~\cite{chakram2021seamless}, as illustrated in Fig.\ref{fig:gates and Setup}a. We use the TE102-Alice and TE202-Bob as memory modes for encoding the cat states, and the TE101 for dispersively reading out the transmon ancilla state. In our design, a single pin-antenna sets the lifetime ($T_1$) of our modes (see Table \ref{table:CohTi}). The pin is located at the nodes of the Alice and Bob memory modes and at the anti-node of the TE101 readout mode. This placement strongly suppresses the coupling of the memory modes to the transmission line while retaining a strong coupling to the readout mode for faster ancilla readout. We achieve almost 2 orders of magnitude difference in the coupling strength, which allows for a simplified setup with a single port, while still providing sufficiently long memory mode lifetime for our experiment~\cite{flurin2017observing}.

\begin{figure}[t!]
    \centering
    {\includegraphics[width=1\linewidth]{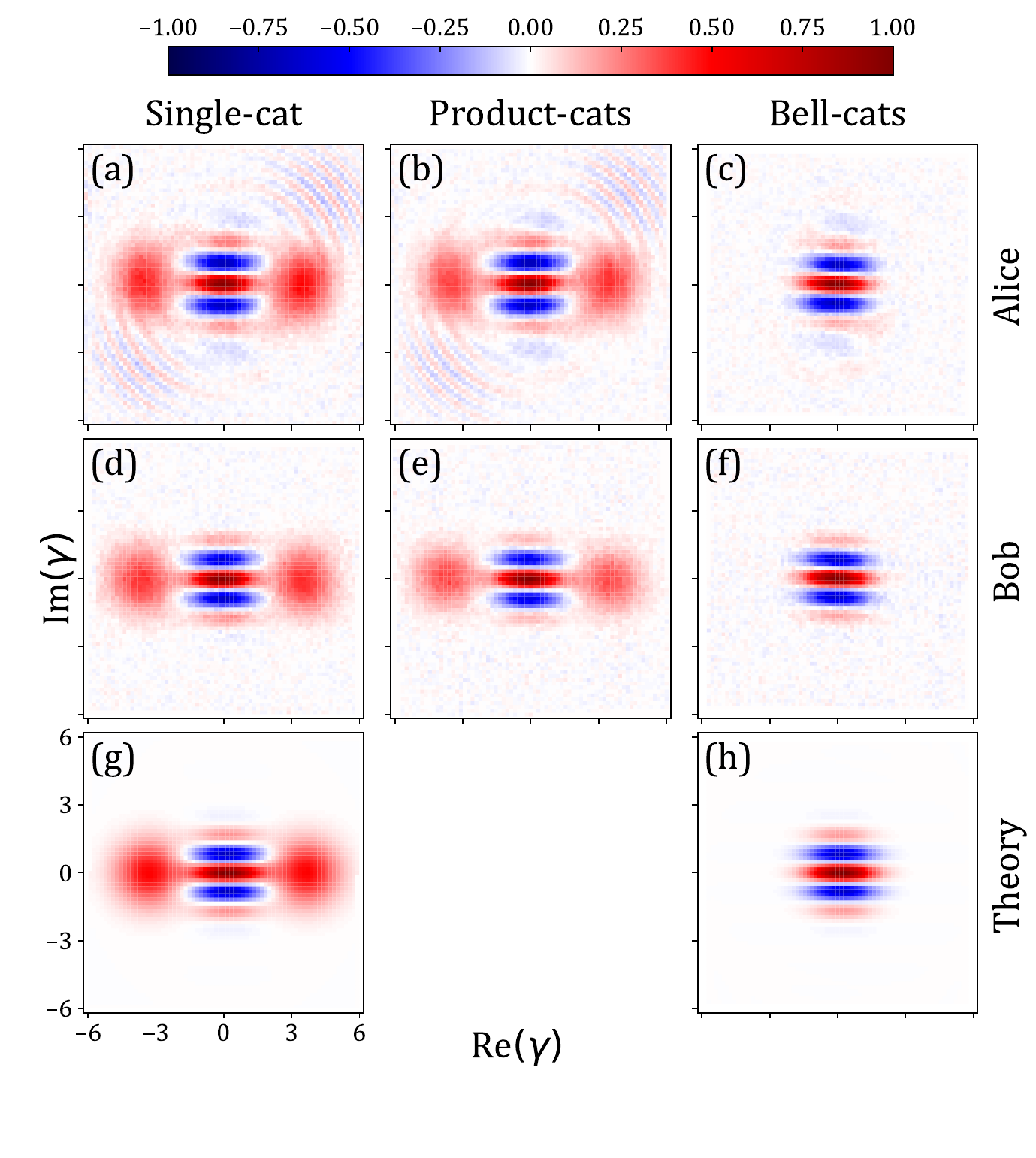}}
    \caption{\textbf{Single-mode characteristic function tomography} Measured real part of the single-mode characteristic functions of Alice $[\mathcal{C}_{A} (\gamma_{A} )]$ (top row) and Bob $[\mathcal{C}_B(\gamma_B)]$ (middle row), and calculated for an ideal state $[\mathcal{C}_{thy} (\gamma_{thy})]$ (bottom row). These are shown for the different cat-states labeled on top. Note that panels (a) and (d) are different states, a single-cat generated in Alice and a single-cat generated in Bob respectively, whereas (b) and (e) are the same generated state, the product-cats state, only measured on Alice and Bob separately. Similarly (c) and (f) are the same state, the Bell-cats state, only measured on Alice and Bob separately. Single-cat and product-cats are expected to be identical in an ideal system where Alice and Bob have completely independent control.  
    }

    \label{fig:single vs simultainous}
\end{figure}

\begin{table}[ht!]

\centering
\begin{tabular}{ l | c c c c c  }
\textbf{Mode}
   &  \boldsymbol{$\omega/2\pi$}  &  \textbf{Mode}  &  \boldsymbol{$T_1$} & \boldsymbol{$T_2$}(Echo)  &  \boldsymbol{$\chi/2\pi$} \\
\hline
\hline
Alice & 6.56 GHz  &  TE102  &  60-62 $\mu s$  &  -  & 216-224 kHz \\
Bob & 8.02 GHz  &  TE202 & 36-38  $\mu s$ & - & 31-36 kHz \\
Ancilla  &  5.37 GHz & - & 12-15  $\mu s$ & 12-14 $\mu s$ & - \\
Readout   & 4.02 GHz & TE101  &  $\sim$ 600 ns & - & $\sim$ 80 kHz\\
\hline
\end{tabular}
\caption{Parameters and coherence times of the EM modes and the transmon ancilla.}
\label{table:CohTi}
\end{table}

To second order our system can be described by the following Hamiltonian~\cite{koch_charge-insensitive_2007},

\begin{equation}
\begin{split}
    \hamil/\hbar = & \   \omega_{A}a^{\dag}a+\omega_{B}b^{\dag}b+\omega_{q}\ketbra ee\\
    - & \left(\chi_{A}a^{\dag}a+\chi_{B}b^{\dag}b\right)\ketbra ee,
\end{split}
\end{equation}
where $\omega_A$, $a^\dagger$ and $\omega_B$, $b^\dagger$ are the Alice and Bob mode frequencies and raising operators respectively, $\omega_q$ is the transmon transition frequency, and $\chi_A$ and $\chi_B$ are the dispersive shifts (cross-Kerr) of each mode with the transmon ancilla. Higher order Kerr coefficients are neglected.

To generate the cat states we use $\mathrm{CNOD}$s and unconditional transmon rotations. In our experiment, the unconditional rotations are implemented in a standard way by applying microwave pulses with Gaussian envelopes. For the CNOD gate, each anti-symmetric pulse is realized using an envelope composed of a multiplication of a Gaussian with a sinusoidal function to achieve anti-symmetry with respect to the carrier frequency $\omega_c^{(e)}$. Typically Gaussian envelopes have several standard deviation widths to make them smooth. We deviate from this and choose a duration that is only twice the standard deviation, resulting in a highly non-smooth drive. This unconventional choice is motivated by the goal of maximizing the accumulated power applied to the harmonic oscillator. As the speed of these gates is limited solely by the maximal driving amplitude, this helps to reduce the duration of our CNOD gates. For the same reason, we select the period of the sinusoidal function to be much larger than the pulse duration, which means that the sine function can be approximated by a linear function. The total duration of a single CNOD operation is determined by the time needed for the unconditional ancilla rotation and the two anti-symmetric pulses. The duration of the ancilla rotation remains constant at 24ns. However, the duration of the anti-symmetric pulses is limited by the available driving amplitude for each mode. During state preparation, the duration of these pulses ranges from $2\sigma$=60ns for the smallest displacements to $2\sigma$=288ns for the largest.

For reconstructing each of the single modes we measure the complex characteristic function Tr$[\rho\mathcal{D}(\gamma)]$~\cite{fluhmann2020direct}. 
For each state, the single-mode characteristic functions of both modes are measured independently, as made possible by the echoing effect of the CNOD. 
Both real and imaginary parts of the characteristic function are analyzed and used to reconstruct the density matrices through maximum likelihood estimation (MLE). This procedure allows us to estimate the fidelities of our states. See supplementary information for exact details, calibrations and procedures.

To quantify the entanglement between the modes we introduce a two-mode joint characteristic tomography by measuring the function, 
\begin{equation}
    \mathcal{C}_J  = \mathrm{Tr}[\rho\mathcal{D}_{\gamma_A}\mathcal{D}_{\gamma_B}],
\end{equation}

where $\mathcal{C}_J$ is the joint, characteristic function~\cite{lai1989characteristic} in the 4D phase space. By applying the CNODs simultaneously on the two modes, as illustrated 
in Fig.\ref{fig:gates and Setup}, our sequence extends the single mode scheme of~\cite{fluhmann2020direct}. The scheme generalizes to multiple modes by introducing N simultaneous or sequential conditional displacements, which allows to measure the N mode characteristic function~\cite{RevModPhys.84.621}. Therefore, non-local correlations can be directly measured using a single qubit ancilla.


We start by generating independent cats. We create cat-states in each mode while keeping the other mode in a vacuum, which we refer to as single-cat. We compare these to a two-mode product state, where a cat-state resides in each mode, which we refer to as product-cats. In the ideal setting, we expect them to be identical. 

Our constructed CNODs implement conditional displacements on each mode which do not depend on the state of the other. The $\pi$-pulse in the middle of the CNOD gate creates an echoing effect. This ensures that while we manipulate one of the modes the other mode does not evolve except for an unconditional phase space rotation, which we negate by a digital rotation. We apply the sequences shown in Fig. \ref{fig:gates and Setup}b. Specifically, we generate the single-cat states in either Alice $\ket{\psi_{\mathrm{s}_{\mathrm{A}}}}=\mathcal{N}(\ket{\alpha}_A + \ket{-\alpha}_A) \otimes \ket{0}_B$ or Bob $\ket{\psi_{\mathrm{s}_\mathrm{B}}}=\mathcal{N}\ket{0}_A \otimes (\ket{\alpha}_B +\ket{-\alpha}_B)$ separately, where $\mathcal{N}$ is a normalization factor. For this we apply the CNODs on the relevant mode only. To generate the product-cats state $\ket{\psi_\mathrm{p}}=\mathcal{N}(\ket{\alpha}_A + \ket{-\alpha}_A) \otimes (\ket{\alpha}_B +\ket{-\alpha}_B)$ we apply the same operations sequentially, first on Bob and then on Alice. 

The real parts of the characteristic functions are displayed in Fig.\ref{fig:single vs simultainous}, the imaginary part is not shown as it is very close to zero everywhere (example is shown in the supplementary materials). The results for the single-cat states and the corresponding parts of the product-cats state look almost identical, as we would expect (and desire) for individual mode control. This is in contrast to previous work where the inherited cross Kerr is substantial (large $\chi$s)~\cite{wang_schrodinger_2016}. One visible discrepancy in our data is the positivity of the single-cat vs the product-cats of Bob. The smaller positivity in the product-cats is due to the sequential creation of the cats which leads to longer operation time. The result is additional decoherence due to photon loss. The second visible discrepancy is a small rotation of a couple of degrees due to the cross-Kerr interaction during the CNOD itself. Importantly, this rotation is independent of the state in either mode, it depends only on the CNOD being applied and can be digitally corrected. This correction has been applied experimentally in the other datasets shown and is discussed in the supplementary materials.  

We now proceed to the entangled cats demonstrating the two-mode control required for universality. When applied to multiple modes conditioned on the same ancilla, the CNOD gate allows to entangle different modes. Combined with unconditional ancilla rotations, the set $\{\mathrm{CNOD}(\vec{\alpha_i}),R(\phi,\theta)\}$, where $i$ is the mode index, is universal in the control of the Hilbert space comprised of multiple Bosonic modes and a single ancilla qubit (see supplementary materials). 
We demonstrate the entangling properties of our gate set by generating the Bell-cats state $\ket{\psi_{\mathrm{b}}}\equiv \mathcal{N}(\ket{\alpha_A, \alpha_B}+\ket{-\alpha_A, -\alpha_B})$ using the sequence illustrated in Fig.\ref{fig:gates and Setup}b with $\alpha_A \simeq \alpha_B \simeq 1.7$. 

Specifically, we implement a two mode CNOD gate  $\mathrm{CNOD}(2\alpha_A,2\alpha_B)$, realizing the three-way entangling operation $\frac{1}{\sqrt{2}}(\ket{g}+\ket{e})\ket{0}_A\ket{0}_B \rightarrow  \mathcal{N}(\ket{e}\ket{\alpha}_A \ket{\alpha}_B - \ket{g} \ket{-\alpha}_A \ket{-\alpha}_B)$ between the Alice and Bob modes and the ancilla. Then an unconditional ancilla rotation followed by a second, small CNOD at the perpendicular axis $\mathrm{CNOD}(i\beta_A,i\beta_B)$ with $\sum_{i =1,2}\alpha_i\beta_i =\frac{\pi}{2}$, is used to disentangle the ancilla. To apply the first two-mode CNOD a single-mode CNOD is applied on Bob and then on Alice. Simultaneous application of the CNOD to both modes is possible and faster, yet we found that it induces undesired effects for large displacements (see supplementary materials). These are attributed to coherent transitions at high photon numbers. It should be possible to circumvent this restriction, yet we postpone such analyses for future work. We do use a simultaneous application of the two-mode CNOD for the second disentangling CNOD gate, which implements small displacements only.

We start with the characteristic functions of each individual mode separately for the Bell-cats state  $\mathcal{C}_i(\gamma_i)=Tr[\mathcal{D}(\gamma_i)\rho]$ (i = A, B), which are shown in the right column of Fig.\ref{fig:single vs simultainous}. We see in the measured single-mode characteristic functions $\mathcal{C_A}$ and $\mathcal{C_B}$ that on their own the state of each mode is in a statistical mixture of two different coherent states, as expected. In the figure we can see the absence of blobs, indicating the expected mixture.

\begin{figure}[t!]
    \centering
    {\includegraphics[width=1\linewidth]{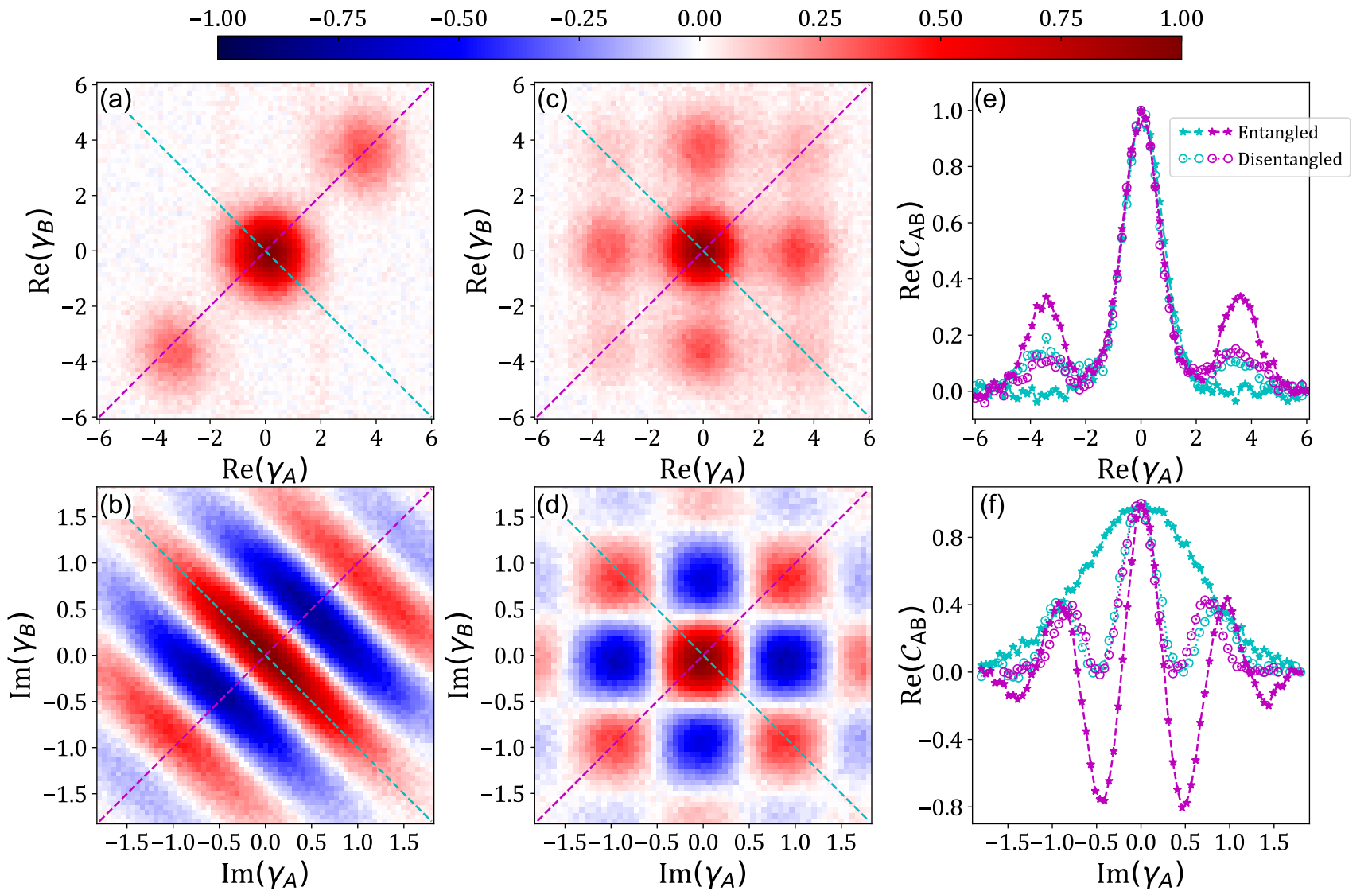}}
    \caption{\textbf{Joint characteristic  tomography of Bell-cats compared with product-cats} 2D cuts of the 4D joint characteristic functions $[\mathcal{C}_{J} (\gamma_{A},\gamma_B )]$ for the Bell-cats (a)-(b), and for product-cats (c)-(d). Cuts are along the planes, Re$(\gamma_A)$-Re$(\gamma_B)$ ((a) and (c)) and Im$(\gamma_A)$-Im$(\gamma_B)$ ((b) and (d)). 1D cuts from the 2D planes corresponding to the dashed lines ((e)-(f)).}
    \label{fig:EW entagled vs simultanious}
\end{figure}

The core features of the Bell-cat state $\ket{\psi_b}$ in the 4D characteristic function are found in 2D cuts along the Re$(\gamma_A)$-Re$(\gamma_B)$ and Im$(\gamma_A)$-Im$(\gamma_B)$ planes. We plot the cuts for an ideal state and for measured data from dataset 1 (Fig.\ref{fig:AS pulse}c). We also plot for comparison, the 2D cuts for the product-cats and adjacent to it measured data from dataset 2 (Fig.\ref{fig:EW entagled vs simultanious}).
There are a few prominent features of the Bell-cats in the 2D cuts, as compared with theory and the product-cats. The Re$(\gamma_A)$-Re$(\gamma_B)$ cut of $\ket{\psi_b}$ contains three positively valued Gaussians spheres (a.k.a blobs). The phase is conveyed by the two blobs centered around $(\pm 2\alpha_A,\pm 2\alpha_B)$, and would be absent for coherent states. For a pure Bell-cats state, the maximum of these blobs should be 0.5, and is $\simeq 0.36$ in dataset 2. The effect is mostly attributed to photon loss which changes the parity of the state.

The fringes in the Im$(\gamma_A)$-Im$(\gamma_B)$ cut indicate the correlation between the coherent states in each of the different modes. The maximum is 0.96 compared with an ideal state which exponentially approaches unity with cat size. For disentangled cats we observe the checkerboard pattern, which is a result of the product of orthogonal fringes from the two separate cats in the product state,  Fig.\ref{fig:EW entagled vs simultanious}d.

Using two-mode CNOD operations we can map on to the ancilla correlations between the Alice and Bob modes and measure them directly. Specifically, we regard each mode as a logical cat-qubit in the standard cat-codes $\ket{0_L / 1_L}\equiv\ket{\alpha / -\alpha}$, such that $\ket{\psi_b}=\mathcal{N}(\ket{0_L0_L}+\ket{1_L,1_L})$. We directly measure expectation values of both single cat-qubits and two cat-qubits logical operators (see supplementary materials for the exact mapping protocol). 
This mapping allows us to extract the entanglement witness of our Bell-cats by measuring 4 points only of $\mathcal{C}_J$ to obtain the direct fidelity estimation~\cite{flammia_direct_2011} (2 are sufficient, yet we use 4 for symmetry). We use dataset 2 shown in Fig.\ref{fig:EW entagled vs simultanious} and obtain $\frac{1}{4}(\langle \hat{I}\hat{I}\rangle +\langle \hat{Z}\hat{Z}\rangle+\langle \hat{X}\hat{X}\rangle-\langle \hat{Y}\hat{Y}\rangle) \approx 84\% $ against the ideal Bell-cat, surpassing the classical correlation bound of 50\%. 

We employed our CNOD operations in a multi-mode cavity to create single-, product-, and Bell-cats states. Our CNOD method allowed fast operation in the weak coupling regime in a multi-mode cavity. The times for generating the different cats (Fig.~\ref{fig:gates and Setup}) give speedups of almost two orders of magnitude as compared with $\frac{3 \pi}{\chi}$ from previous work~\cite{wang_schrodinger_2016}. In practice we were limited by the available amplifiers, further optimization may yield faster operations. while the anti-symmetric shape should make the CNOD rather robust. We can also make a comparison between the CNOD method and the ECD gate, which was demonstrated on a single mode~\cite{eickbusch2022fast} (see supplementary materials). In the ECD method, fast conditional displacements are achieved using short oscillator drives approximated as delta functions, with a wait time separating them. In contrast, our CNOD method utilizes anti-symmetric pulses that require less peak power. Additionally, we observed the absence of increased transmon decoherence at large intermediate photon numbers, which is typically expected from higher-order transitions as observed in the ECD experiment~\cite{reed2010high,sank2016measurement, ShillitoPhysRevApplied.18.034031}.. While the underlying cause for this discrepancy is uncertain and may be attributed to differences in experimental setups and specific details, it is possible that the diabatic nature of Landau-Zener transitions in the CNOD scheme helps suppress this effect [29]. A detailed analysis, comprehensive comparisons, and further investigation of the speed limit will be the focus of future work.

The combination of weak coupling and fast operation enabled by the CNOD method was the key to the high-quality state generation in a multi-mode setting, despite the rather pedestrian cavity photon lifetimes. The photon lifetimes are the main contributor to the infidelity and can be improved by a lot, these were set by the pin-antenna coupling. Using a single pin-antenna for both memory (input) and readout (output) modes serves as a double-sided sword. It simplifies the setup, however, in practice, there is a limit on the coupling ratio of readout to memory modes. Increasing the ratio far beyond the current 2 orders of magnitude may prove rather challenging in this setup. Sacrificing the setup simplicity and separating the readout and memory can give more than an order of magnitude lifetime improvement~\cite{chakram2022multimode}.

Our findings serve as a proof-of-principle demonstration that a single ancilla weakly coupled to a multi-mode system is sufficient for achieving universal control, offering an efficient method for controlling modules with few modes. Additionally, we show that the cross-talk between these modes can be significantly reduced without compromising the gate speed, thanks to the implementation of our CNOD method.
This creates a path to efficient multi-mode continuous variable encoding~\cite{royer2022encoding}. Extending beyond a few modes could be achieved by coupling such modules through a control bus and additional coupling elements~\cite{zhou2021modular}. Overall our approach can help reduce hardware overhead for Bosonic encoding.

\section*{Acknowledgments}
This research was supported by the Israeli Science Foundation (ISF), Pazi foundation, and Technion's Helen Diller Quantum Center. The work made use of the Micro Nano Fabrication Unit at the Technion. L.J. acknowledges support from the ARO MURI (W911NF-21-1-0325), AFOSR MURI (FA9550-19-1-0399, FA9550-21-1-0209), NSF (OMA-1936118, ERC-1941583, OMA-2137642), NTT Research, and the Packard Foundation (2020-71479). We thank A, Eickbush, A. Turner, B. Katzir, I. Kaminer, G. Moshel, and D. Oren Caspi for useful discussions.

\appendix

\renewcommand{\thesection}{\Alph{section}}

\section{Experimental setup}\label{Setup}
\begin{figure}[htp!]
    \centering
    {\includegraphics[width=1\linewidth]{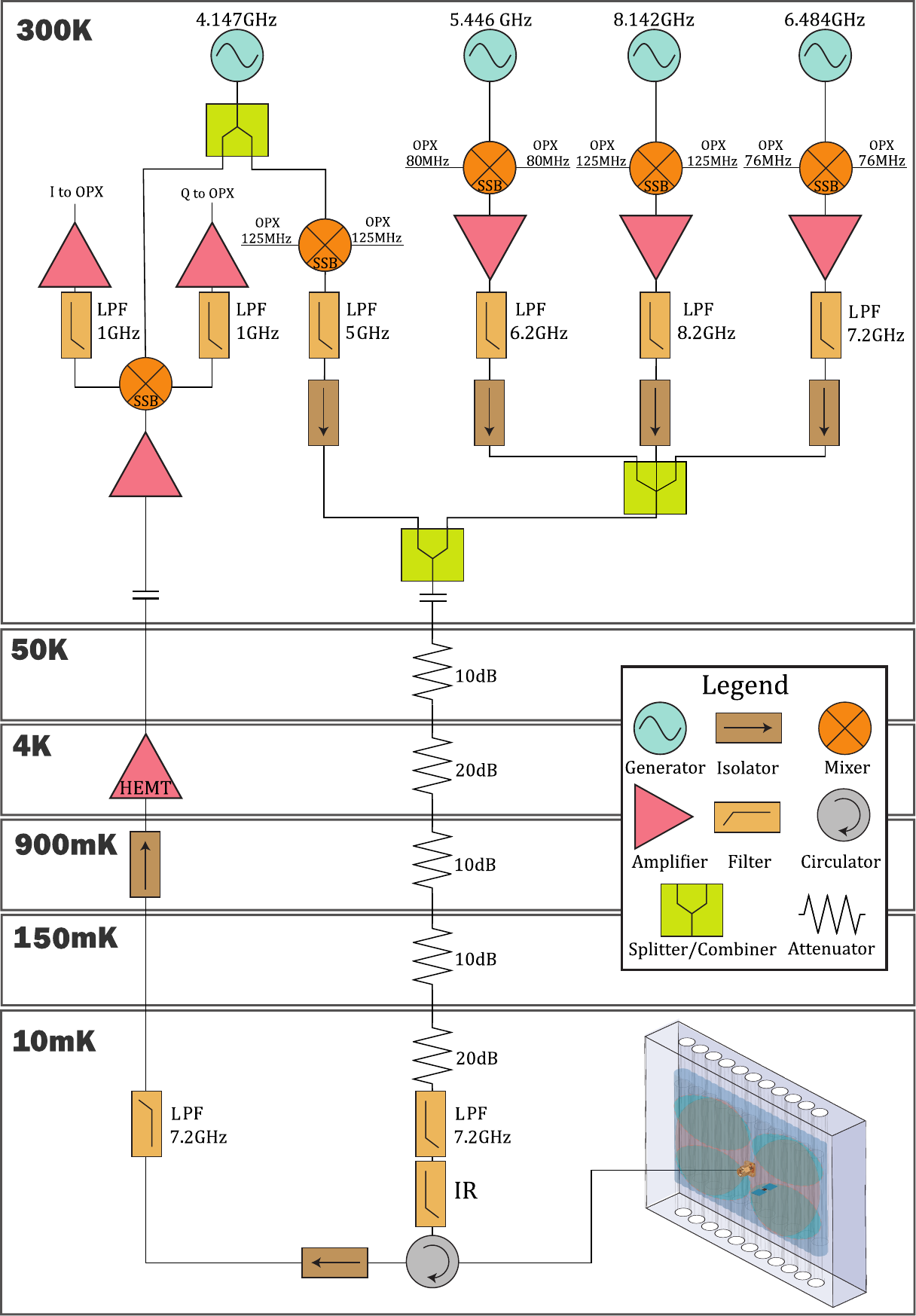}}
    \caption{\textbf{Schematic illustration of the experimental setup.} The flute cavity and the coupled transmon ancilla are placed at the bottom stage of a dilution refrigerator and shielded by a Cryoperm-Copper-Tin shield. We use an Operator-X (OPX) by Quantum Machines, to generate low frequency pulses with arbitrary shapes. We mix the low frequency signals with high frequency signals from our generators. We send all the control signals, for the transmon and the memory modes, and the readout signals for the readout mode, down a single attenuated line. The readout signal is reflected through the flute cavity, amplified, and digitized at room temperature.}
    \label{fig:fullSystem}
\end{figure}
The experimental system wiring is schematically represented in Fig.~\ref{fig:fullSystem}. The readout and the Alice and Bob storage oscillators are modes of the microwave flute cavity \cite{PhysRevLett.127.107701} machined out of a single slab of high-purity N5 Aluminum. Following Ref.~\cite{zhao2009millisecond}, the seamless cavity was then etched in Transene A Aluminum etchant for 4 hours.

As a two level ancilla, we use the two lowest levels of a transmon which was fabricated by Aluminum deposition on resist patterns formed by electron beam lithography, with a layer of ZEON ZEP 520A resist on a layer of MicroChem 8.5 MMA EL11 resist on top of a silicon substrate. Development of the resist was done at room temperature for MMA and at 0C for ZEP. The Al/AlOx/Al Josephson junction was fabricated using a bridge-free process~\cite{zhang2017bridge}. 

To couple the ancilla to the oscillators, the edge of the silicone chip opposite to the transmon is clamped as the superconducting circuit is inserted through a 12mm tunnel of with diameter of 4mm into the cavity. 

\section{Full Hamiltonian and parameters}\label{Full-Hamil}
During the application the CNOD gate, our system reaches large photon numbers. Therefore, the Hamiltonian providing an effective description of our undriven system should include higher order terms and is given by 
\begin{equation}
\begin{split}
    \hamil/\hbar = & \	\omega_{A}a^{\dag}a+\omega_{B}b^{\dag}b+\omega_{q}\ketbra ee+\omega_{R}r^{\dag}r\\
    - & \left( 	\chi_{A}a^{\dag}a+\chi_{B}b^{\dag}b-\chi_{R}r^{\dag}r\right) \ketbra ee\\
    + & 	\frac{K_{A}}{2}a^{\dag}a^{\dag}aa-\frac{K_{B}}{2}b^{\dag}b^{\dag}bb-K_{AB}a^{\dag}ab^{\dag}b \\
    - &  \left(\frac{K_{R}}{2} r^{\dag}r+K_{AR}a^{\dag}a+K_{BR}b^{\dag}b \right)r^{\dag}r,
\end{split}
\end{equation}
where $a,b$ and $r$ are the bosonic annihilation operators of the oscillator like Alice, Bob and readout mode respectively and $\ketbra{e}{e}$ is the excited state of the transmon ancilla which is treated as a two-level system. 
\begin{table}[ht!]

\centering
\begin{tabular}{ l | c c c  }
\textbf{}
   &  \boldsymbol{$K_{Ai}/2\pi$}  &   \boldsymbol{$K_{Bi}/2\pi$}   &   \boldsymbol{$K_{Ri}/2\pi$} \\
\hline
\hline
Alice & 125 Hz  &  19 Hz    & 45 Hz \\
Bob & 19 Hz  &  3 Hz & 7 Hz    \\
Readout   &  45 Hz & 7 Hz &  16 Hz \\
\hline
\end{tabular}
\caption{Estimated self ($K_{i}$) and cross ($K_{ij}$ Kerr coefficients of the EM mode calculated as $K_{ij}=\frac{\chi_i\chi_j}{2K_t}$ where $K_t\approx 194$MHz is the transmon anharmonicity and $\chi_i\equiv K_{ti}$ is the dispersive coupling between the transmon and the mode $i$~\cite{nigg2012black}.}
\label{table:KerrCoffeiceints}
\end{table}

\section{A single anti-symmetric pulse}\label{sec:SinglePulse}
\noindent
\begin{figure*}[t]
\includegraphics[scale=0.5]{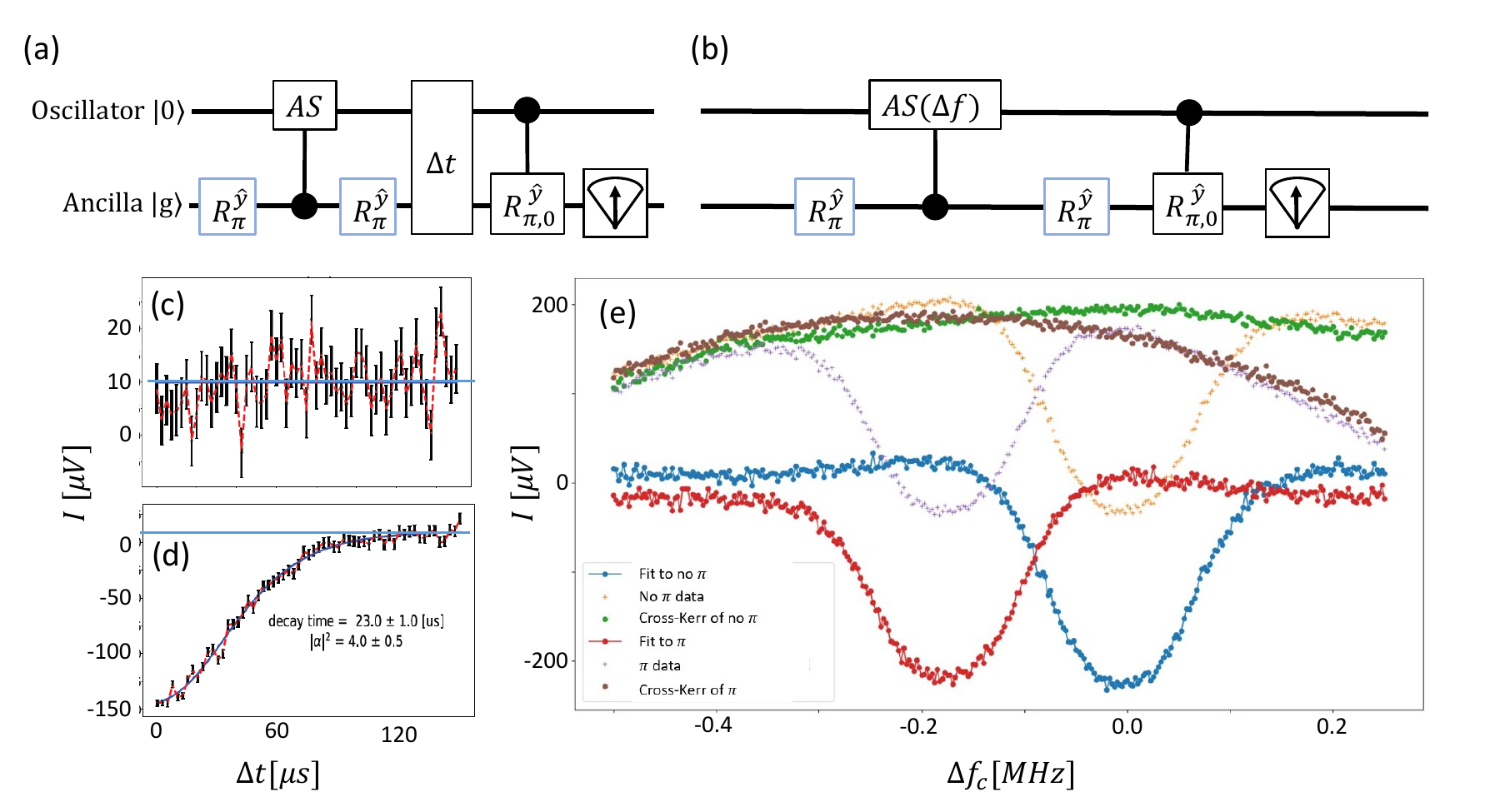}
\centering\caption[Conditionality of an anti-symmetric pulse]{\textbf{Single anti-symmetric pulse.}
A modified version of the protocol from~\cite{reagor2016quantum} (a), is used here for demonstrating the conditionality of the anti-symmetric pulse. Cavity displacement as function of wait time $\Delta t$ is shown for when the transmon is either in in-ground state (d) or excited state (c) during the anti-symmetric pulse. Transmon state is determined by an unconditional $\pi$ rotation (teal). The blue horizontal line in both panels marks the time-averaged value of the results displayed in panel (c).
In either case, the transmon is returned to the ground state prior to the application of the narrow-band conditional $\pi$ rotation. The cross-Kerr between the readout and memory modes is accounted for through a control experiment differing from the original by the absence of the final conditional rotation performed at each stage. 
Protocol for $\pi$ no-$\pi$ spectroscopy (b), which is similar to (a) yet with a zero wait time ($\Delta t$) and varying the detunings ($\Delta f_c$) of the frequency of the node of the anti-symmetric pulse. Spectroscopic measurements (e) show voltage of the readout mode as function of $\Delta f$ for no-$\pi$,$\pi$ (purple,orange) and for the cross-Kerr control experiment (brown, green). Subtracting the control data gives values proportional to the occupation probability of the transmon excited state (red, blue). The minima of each plot indicate the maximal excitation of the transmon, corresponding to the minimal (vacuum/thermal state) of the memory mode after the anti-symmetric pulse. As each minima corresponds to the appropriate state-dependent frequency of the memory mode, the coupling strength $\chi$ can be deduced by subtracting the two.}

\label{fig:As:Condtionality}

\end{figure*}
For realizing a conditional displacement, the full CNOD scheme is required. However, the strong conditionality is present already at the level of a single anti-symmetric pulse. In contrast to the full scheme, a single anti-symmetric pulse displaces the oscillator only if the ancilla is in one state, while it rotates if the ancilla is in the other. Therefore, for the single pulse, the average photon number, and thus the energy of the oscillator changes or not dependent on the state of the ancilla, this is in contrast to the ECD method~\cite{eickbusch2022fast}.

One way to exploit this behavior is by utilizing schemes that involve a single anti-symmetric pulse as the sole drive applied to an EM mode. For such schemes, the oscillator should initially be in the rotation-symmetric vacuum or thermal states. This simplified state configuration facilitates the analysis by rendering the conditional phase space rotation inconsequential, thus it can be ignored. Further simplification is achieved by constraining the ancilla initial state to its eigenstates. When the anti-symmetric pulse is applied to this initial state, the memory mode either remains in the vacuum state if the pulse node aligns with the frequency of the EM mode corresponding to the ancilla state or undergoes displacement otherwise. Thus, we can investigate attributes of the anti-symmetric pulse, as well as aspects of the experimental setup, by effectively mapping the EM mode's vacuum state to the ancilla's state through a conditional (slow) $\pi$ rotation.

One example of such a scheme is the sequence presented in Fig.~\ref{fig:As:Condtionality}a, whose experimental results are displayed in Fig.~\ref{fig:As:Condtionality}c,d (We note that the data displayed here is taken on a different, yet similar setup than the one in the main text) . The sequence is utilized, either with two unconditional $\pi$ pulses (Fig.~\ref{fig:As:Condtionality}c) or without (Fig.~\ref{fig:As:Condtionality}d). When applying the $\pi$ pulses we find a constant value, indicating that the memory mode returns to the vacuum state if the transmon is excited during the anti-symmetric drive. However, when the transmon remains in the ground state during the drive, the quadrature displays dynamics. For short wait times ($\Delta t$) there is a displacement which decays and eventually reaches values corresponding to vacuum. This procedure highlights that the evolution of the memory mode induced by the anti-symmetric pulse is highly distinguishable for each transmon state. One way to take advantage of this behaviour, which is inherently different from that of the CNOD, is by probing the experimental system through a $\pi$ no-$\pi$ spectroscopy procedure, Fig.~\ref{fig:As:Condtionality}b.

The $\pi$ no-$\pi$ spectroscopy applies the previously mentioned procedure with a zero wait time ($\Delta t$), varying the detunings of the frequency of the node of the anti-symmetric pulse.

Intuitively, following the anti-symmetric pulse, the memory mode is \textit{not} displaced if its frequency corresponds to that of the pulse node. Therefore, by fitting each of the results, we can find the corresponding cavity frequency expected at the maximal vacuum occupation. By comparing the $\pi$ and no-$\pi$ results, we can evaluate $\chi$.

The advantage of the $\pi$ no-$\pi$ spectroscopy over other memory mode characterization schemes lies in the intermediate conditional operation being fast enough to avoid the typically short transmon coherence times. Furthermore, as the displacement of the memory mode by the conditional displacement can be rather large, the selectivity of the final conditional $\pi$ rotation can be relaxed.

Another possible application of the $\pi$ no-$\pi$ spectroscopic method is for measuring the higher order terms of the setup. Such terms as the self-Kerr $K_c a^{\dagger 2} a^2 $, which are absent from the dispersive Hamiltonian, become increasingly important at large photon numbers. Therefore, while the effects of these terms are usually negligible, they can have a significant impact as, during the application of the anti-symmetric pulse, the trajectory of the EM mode occupies large photon states. For example, in the presence of the anti-symmetric drive, the effective frequency of the memory mode can shift by an amount proportional to the self-Kerr coefficient $K_c$ and the intermediate photon numbers, whose average value is dominated by the anti-symmetric pulse itself.

Furthermore, as previously noted, the magnitude of the final displacement by an anti-symmetric pulse is linear in the driving amplitude but squared in the pulse duration. Therefore, we can use such scaling to reach similar final states by traversing different trajectories of the memory modes' phase space. Studying such trajectories by sweeping not only frequency but also driving amplitude and/or pulse duration can be useful for evaluating the higher order terms, studying qubit ionization, and even optimizing the $\mathrm{CNOD}$ gate.
 
\section{Ancilla to cat mapping}\label{QC-mapping}
We show the mapping of an ancilla state on to the Bloch sphere of single-mode cat code, which is generalized to multiple modes. For an initial state,
\begin{align}
 \ket{\psi_i}=\left(\cos\left(\frac{\theta}{2}\right)\ket {g}-i\sin\left(\frac{\theta}{2}\right)e^{i\phi}\ket e\right)\otimes\ket{\vec{0}} 
\end{align}
where $\vec{0}$ is the vacuum of all modes. Applying the $\mathrm
{CNOD}\left(\Vec{\alpha}\right)$ gate followed by a $\mathcal{R}_{\hat{y}}(\pi/2)$ rotation, up to global phase, we get,

\begin{align}
 &\frac{\cos\left(\frac{\theta}{2}\right)}{\sqrt{2}}\left(\ket{e,\frac{\vec{\alpha}}{2}}-\ket{g,\frac{\vec{\alpha}}{2}}\right) \nonumber \\
+i&\frac{\sin\left(\frac{\theta}{2}\right)}{\sqrt{2}}e^{i\phi}\left(\ket{g,-\frac{\vec{\alpha}}{2}}+\ket{e,-\frac{\vec{\alpha}}{2}}\right)
\end{align}

To disentangle the modes from the ancilla we perform a second CNOD gate with $i\vec{\beta}$, 

\begin{align}
\begin{split}
&\frac{\cos\left(\frac{\theta}{2}\right)}{\sqrt{2}}\left(\ket{g,\frac{\vec{\alpha}}{2}-i\frac{\vec{\beta}}{2}}e^{-i\vec{\alpha}\cdot\vec{\beta}/2}+\ket{e,\frac{\vec{\alpha}}{2}+i\frac{\vec{\beta}}{2}}\right) \\
+&i\frac{\sin\left(\frac{\theta}{2}\right)}{\sqrt{2}}e^{i\phi}\left(\ket{g,-\frac{\vec{\alpha}}{2}-i\frac{\vec{\beta}}{2}}-\ket{e,-\frac{\vec{\alpha}}{2}+i\frac{\vec{\beta}}{2}}e^{-i\vec{\alpha}\cdot\vec{\beta}/2}\right)
\end{split}
\label{eq:DisintStart}
\end{align}
where we have once again omitted a global phase.

For $|\vec{\beta}|<<1$, the state can be approximated  as,

\begin{align}
&\approx \frac{\cos\left(\frac{\theta}{2}\right)}{\sqrt{2}}\left(\ket{g,\frac{\vec{\alpha}}{2}}e^{-3i\vec{\alpha}\cdot\vec{\beta}/4}+\ket{e,\frac{\vec{\alpha}}{2}}e^{i\vec{\alpha}\cdot\vec{\beta}/4}\right) \nonumber \\
&+i\frac{\sin\left(\frac{\theta}{2}\right)}{\sqrt{2}}e^{i\phi}\left(\ket{g,-\frac{\vec{\alpha}}{2}}e^{i\vec{\alpha}\cdot\vec{\beta}/4}-\ket{e,-\frac{\vec{\alpha}}{2}}e^{-i3\vec{\alpha}\cdot\vec{\beta}/4}\right) \label{eq:cat_approx}
\end{align}
where we used  $\braket{\delta}{\eta}=e^{iIm\left(\delta^{*}\eta\right)}e^{-\frac{1}{2}\left|\delta-\eta\right|^{2}}\approx e^{iIm\left(\delta\eta^{*}\right)}$. For  $\vec{\beta}\cdot\vec{\alpha}=\frac{\pi}{2}$, up to a global phase we get,
\begin{equation}
\begin{split}
\ket{+i}\otimes\left(\cos\left(\frac{\theta}{2}\right)\ket{\frac{\vec{\alpha}}{2}}-\sin\left(\frac{\theta}{2}\right)e^{i\phi}\ket{-\frac{\vec{\alpha}}{2}}\right)
\end{split}
\end{equation}

where $\ket{+i}=\frac{\ket g+i\ket e}{\sqrt{2}}$.

\begin{figure}[htp!]
    \centering
    \includegraphics[width=1\linewidth]{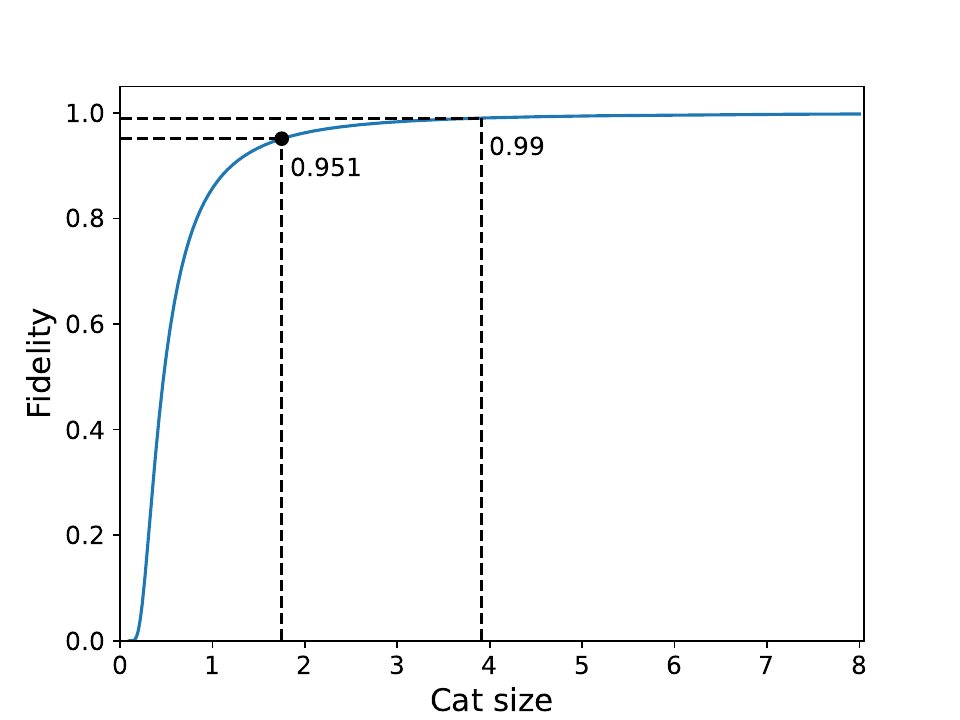}
    \caption{\textbf{Mapping fidelity as a function of cat size} To assess the fidelity of our mapping, taking into account the potential limitations of the approximation in equation~\ref{eq:cat_approx}, we calculate the fidelity between the final state generated by the ideal unitary operations on vacuum $U_{+}\ket{0}$  and the target state $\ket{\psi_{targ}}\propto \ket{+i}\otimes\left(\ket{\frac{\vec{\alpha}}{2}}+\ket{-\frac{\vec{\alpha}}{2}}\right)$ as a function of the cat size $\alpha$. As the cat size increases, the fidelity also increases, indicating improved qubit-cavity disentanglement. The marked point on the graph represents the fidelity corresponding to the cat sizes generated in our experiment. We observe that a $99\%$ fidelity is achieved at a cat size twice as large.}
    \label{fig:disenterror}
\end{figure}
\section{Calibrations of pulses and cat phase}\label{Calibrations}
In this section, we describe the different methods that we used to calibrate the CNOD amplitude and acquired geometric phase, and to measure the parameters of the system. 

\begin{figure}[htp!]
    \centering
    {\includegraphics[width=1\linewidth]{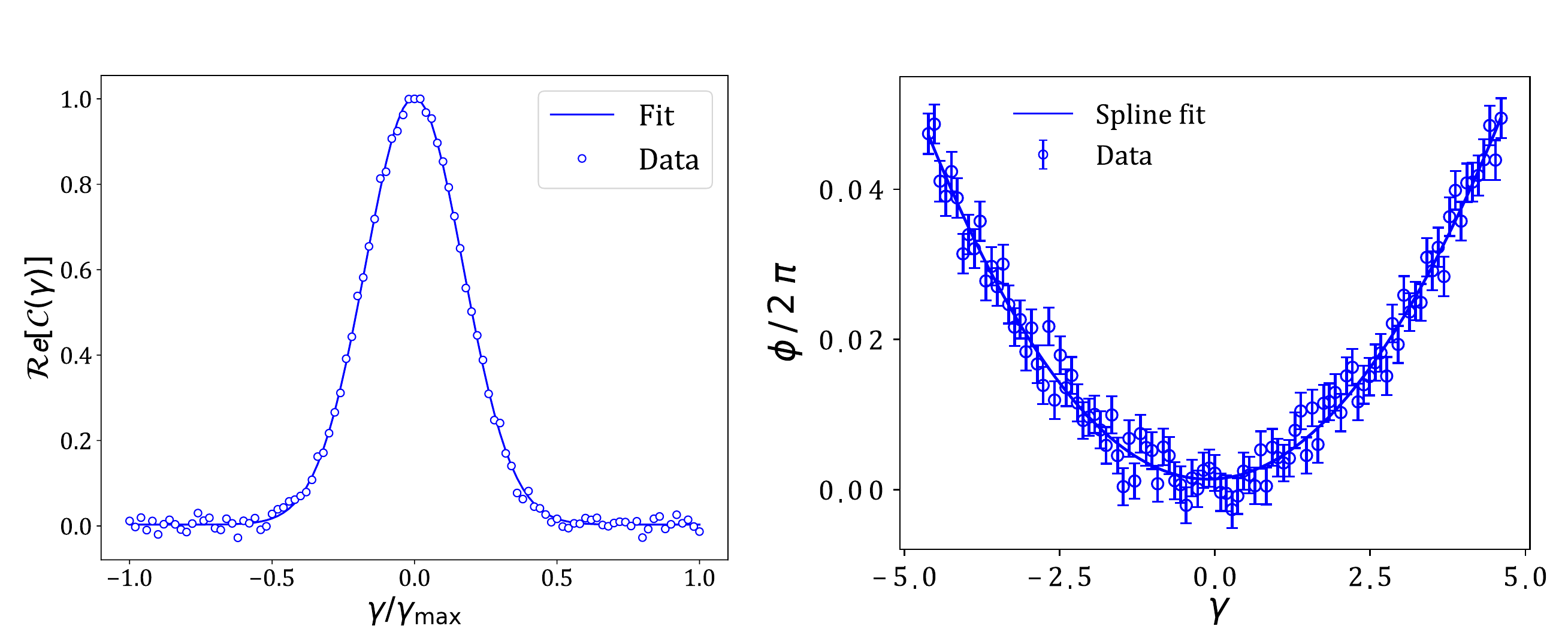} }
    \caption{\textbf{Calibration of CNOD amplitude and acquired geometric phase.} (a) The characteristic function of a vacuum state along an arbitrary axis of length $2\gamma_{\mathrm{max}}$. By fitting the data to the expected Gaussian of variance $2$, we calibrate the maximal displacement amplitude $\gamma_{\mathrm{max}}$. (b) The acquired geometric phase as function of the displacement amplitude $\gamma$. We scale the displacement amplitude only by scaling the anti-symmetric pulse amplitude while keeping the pulse time fixed. We use a spline fit to extrapolate the phase for any magnitude of $\gamma$.} 
    \label{fig:pulseCalibration}
\end{figure}

We calibrated the CNOD amplitude by measuring a single axis (1D sweep) of the characteristic function of a vacuum state (example shown in Fig.~\ref{fig:pulseCalibration}a). We fit the data to the expected Gaussian shape, and thus we directly get the amplitude scale. 

A finite thermal population inside each mode would result in an inaccurate calibration, as the Gaussian at the origin of the characteristic function of a thermal state is narrower than that of a vacuum state. When we performed the procedure MLE and reconstructed the density matrix of the different prepared states, we found the inaccuracy in pulse amplitude to be small. If we were to take a finite thermal state the calculated fidelities would be a bit higher due to the narrowing effect of finite temperature. Our results are obtained by assuming a vacuum state, thus the fidelities that we show here are a lower bound of the true fidelities.

We calibrated the geometric phase that is acquired due to a CNOD by performing the following sequence: $\mathcal{R}_{\hat{y}}(\pi/2)$, CNOD$(\gamma)$, CNOD$(-\gamma)$, and
either $\mathcal{R}_{\hat{y}}(\pi/2)$ or $\mathcal{R}_{\hat{x}}(\pi/2)$ followed by a projective measurement of $\sigma_z$ of the transmon ancilla. The two CNODs of opposite directions cancel each other out, and leave the memory mode in the vacuum state, but still impart a geometric phase on the transmon ancilla. By measuring both $\sigma_x$ and $\sigma_y$, we are able to extract the acquired phase, as shown in Fig.~\ref{fig:pulseCalibration}b. 

To calibrate the disentangling pulse amplitude, we sweep over $\beta$, the amplitude of the disentangling CNOD$(\mathrm{i}\beta)$, and measure the purity of the transmon ancilla. For this procedure we measure $\sigma_z$, $\sigma_x$ and $\sigma_y$, but in practice the information about the purity lies in the $x$-axis alone. At first, we pick the amplitude that corresponds to the highest purity. Then we use it to create a cat-state and measure its characteristic function along the displaced axis. We fine tune $\beta$ to minimize asymmetry around $0$, thus maximizing disentanglement.
\begin{figure}[htp!]
    \centering
    \includegraphics[width=1\linewidth]{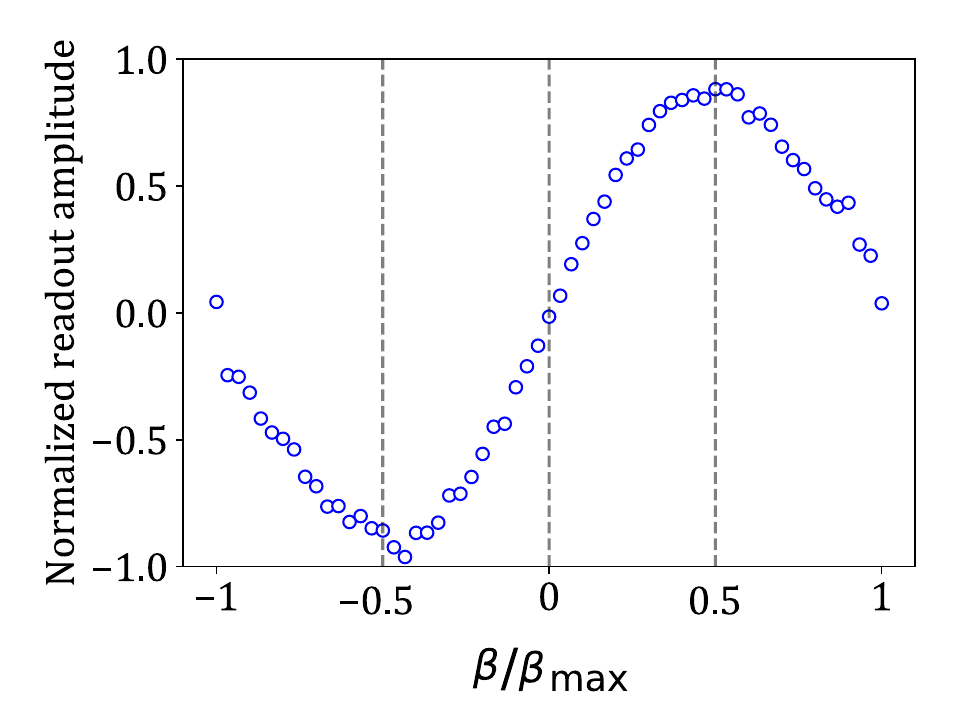}
    \caption{\textbf{Calibration of CNOD amplitude used for disentanglement of the cat.} The expectation value of $\sigma_x$ of the transmon ancilla, that is proportional to the readout amplitude, after creation of a cat. The maximal value of $|\langle\sigma_x\rangle|$ indicates a maximum of purity and therefore corresponds to the ideal disentangling displacement amplitude.} 
    \label{fig:disentCalib}
\end{figure}

\begin{figure}[htp!]
    \centering
    \includegraphics[width=1\linewidth]{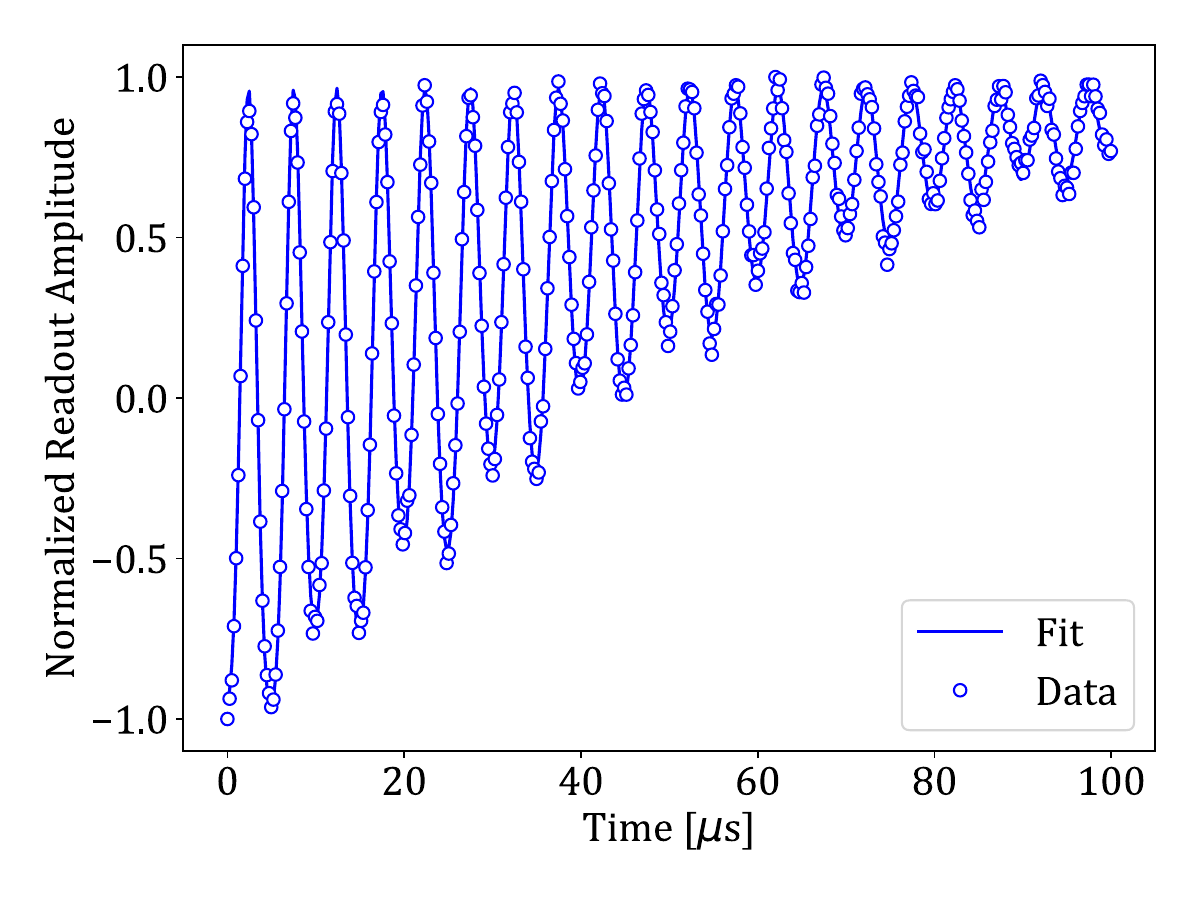}
    \caption{\textbf{Characteristic function of a decaying coherent state.} $\mathcal{C}\left(e^{\mathrm{i}\delta\omega t}\gamma\right)$ versus the evolution time, starting with coherent state $\Ket{\gamma}$. The oscillations are induced by both artificial detuning and detuning between the rotating frame and the resonance frequency of the mode. The decay is dominated by the single photon lifetime of the mode. We fit the data to eq.~\ref{eq:charFuncDecay}.}
    \label{fig:chicalib}
\end{figure}

We accurately measure both the frequency and the lifetime of each mode using a sequence that is similar to Ramsey interferometry. The sequence is composed of $\mathcal{D}(\alpha_0)$, wait a time of $t$, $\mathcal{R}_{\hat{y}}(\pi/2)$, CNOD$(-\mathrm{Exp}(\mathrm{i}\delta\omega t)\gamma)$ and $\mathcal{R}_{\hat{y}}(-\pi/2)$ followed by a projective measurement of $\sigma_z$ of the transmon ancilla (similar to what that was done by~\cite{campagne2020quantum}). We introduce $\delta\omega$ to add artificial detuning and allow for more precise measurement of frequency. The results of one such measurement are shown in Fig.~\ref{fig:chicalib}. As the coherent state $\ket{\alpha(t)}$ rotates and decays (assuming negligible dephasing) the measurement data is expected to behave according to
\begin{equation}
 Re\left[\mathcal{C}_{\ket{\alpha(t)}},\left(e^{\mathrm{i}\Delta\omega t}\beta\right)\right] =A \cos\left(B e^{-t/T^c _1}\sin\left(\left(\omega_d+\Delta\omega\right) t\right))\right),   \label{eq:charFuncDecay}
\end{equation}
where $A=e^{-|\gamma/2|^2}$ and $B=\alpha_0\gamma/2$ are treated as constant scaling coefficients, $T^c _1$ is the mode's single photon life-time, $\Delta\omega$ is the detuning between the rotating frame and the mode's resonant frequency.  In addition, this procedure enables precise measurements of the cross-Kerr between the memory modes, and between each mode and the transmon ancilla. For example, performing this procedure with the transmon at the excited state would decrease $\Delta\Omega$ by $\chi$.
\section{characteristic function tomography}\label{CF-TOMO}
To reconstruct the prepared state in Alice and Bob, we performed characteristic function tomography. We measured both the single-mode characteristic function for each mode and 2D cuts of the 4D joint-characteristic function.

The measurement procedure that we used is composed of a sequence of $\mathcal{R}_{\hat{y}}(\pi/2)$, CNOD$^{(A)}(\gamma_A)$, CNOD$^{(B)}(\gamma_B)$, and 
either $\mathcal{R}_{\hat{y}}(\pi/2)$ or $\mathcal{R}_{\hat{x}}(\pi/2)$ followed by a projective measurement of $\sigma_z$ of the transmon ancilla. This sequence allows us to measure the real part or the imaginary part, respectively, of $\mathcal{C}(\gamma_A,\gamma_B)$. We used CNODs for the measurement procedure longer than for the state preparation in order to reduce the effect of higher order terms such as self- and cross-Kerr on the measurement outcome. We included rotations in two directions that map $\mathcal{C}(\gamma_A,\gamma_B)$ to $-\sigma_z$, and summed the results with the appropriate sign. The expectation value for each point was averaged out of 4000 experiment runs.

\begin{figure}[htp!]
    \centering
    \includegraphics[width=1\linewidth]{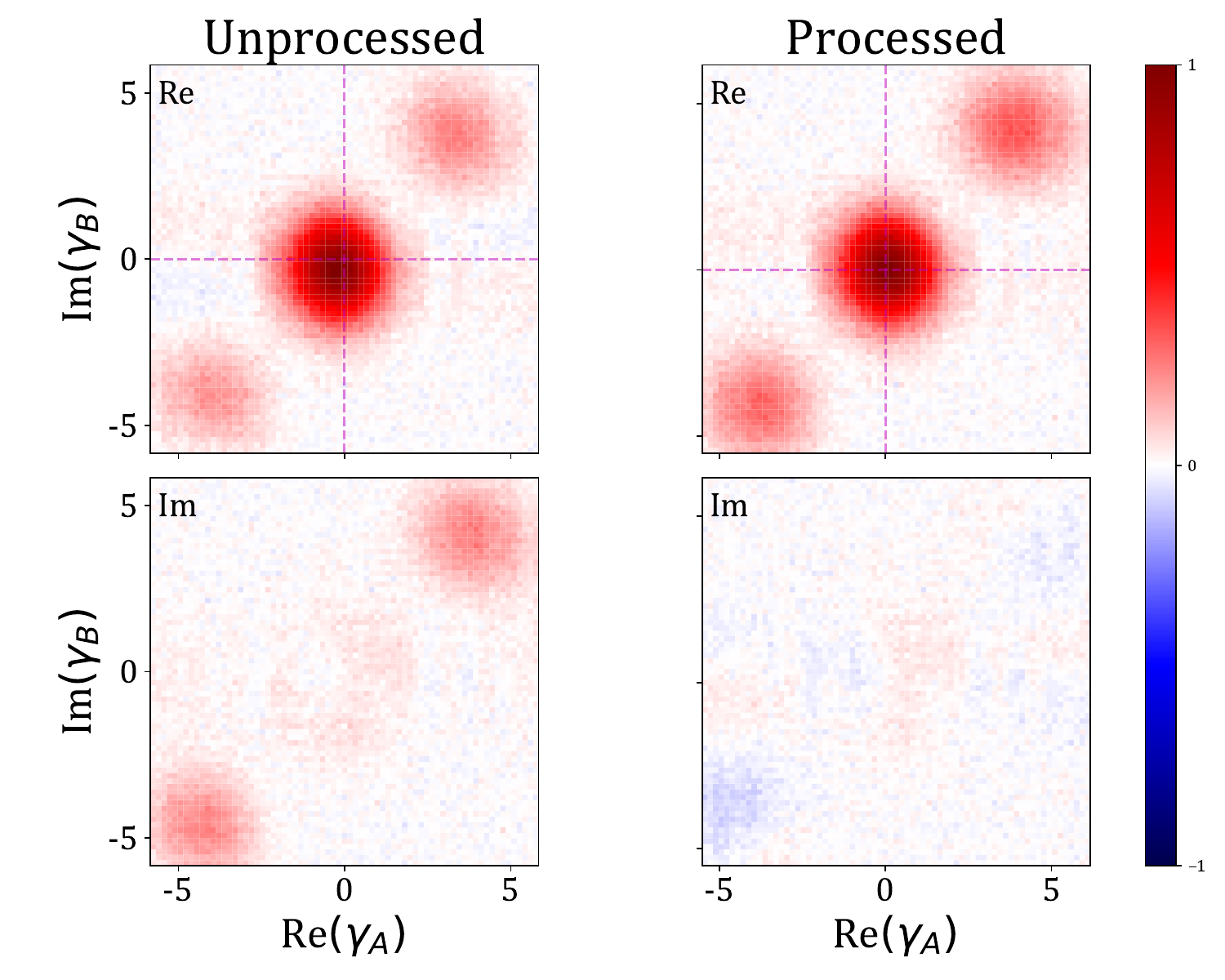}
    \caption{\textbf{Post processing of a joint characteristic function tomography.} The shift in $\gamma_A$ and $\gamma_B$ due to the imperfect disentanglement is corrected as indicated by the dashed lines in the real part of $\mathcal{C}(\gamma)$. The two symmetric blobs in the imaginary part are corrected by accounting for the geometric phase that is acquired by the CNOD. As a result, the imaginary part of the processed data almost vanishes.} 
    \label{fig:postProcessing}
\end{figure}

We apply a couple of required actions in our post-processing procedure. One, the acquired geometric phase due to the CNOD, which was calibrated prior to running the experiment (see section~\ref{Calibrations}), and two the characteristic function was digitally rotated accordingly. In addition, we shifted both $\gamma_A$ and $\gamma_B$ by a real constant, this is to account for the imperfect disentanglement (see section~\ref{sec:imperfect}). An example of this post processing procedure is shown in Fig.~\ref{fig:postProcessing}.

\subsection{Characterstic function with imperfect disentanglement}\label{sec:imperfect}
The disentanglement procedure is imperfect and improves exponentially in cat size. In this work we used relatively smaller cats, and so we calculated the error due to the imperfect disentanglement. The prominent effect is a shift in the characteristic function expected from the experiment, which is accounted for in the post processing. The exact state of the system after the disentangling pulse is (Eq.~\ref{eq:DisintStart} in section~\ref{QC-mapping}),

\begin{align}
&\frac{1}{\sqrt{2}}\bigg(\cos\left(\frac{\theta}{2}\right)\left(\ket{g,\frac{\alpha}{2}-i\frac{\beta}{2}}e^{-i\alpha\beta/2}+\ket{e,\frac{\alpha}{2}+i\frac{\beta}{2}}\right) \nonumber \\
&+i\sin\left(\frac{\theta}{2}\right)e^{i\phi}\left(\ket{g,-\frac{\alpha}{2}-i\frac{\beta}{2}}-\ket{e,-\frac{\alpha}{2}+i\frac{\beta}{2}}e^{-i\alpha\beta/2}\right)\bigg)
\end{align}

If the disentanglement was perfect then after a $\pi /2$ rotation around the X-axis we would end up in the product state of a cat with the qubit in the ground state. Then to measure the characteristic function we would apply the regular procedure, starting with a $\pi /2$ rotation around the Y-axis. Since the disentanglement operation is not perfect, after this procedure we remain with small residual entanglement. Thus, following the ancilla rotations, the state is,

\begin{equation}
\left(\ket{-i}\otimes\ket{\psi_{-i}}-\ket{+i}\otimes\ket{\psi_{+i}}\right)
\end{equation}
where
\begin{align}
\ket{\psi_{-i}}&=\frac{1}{\sqrt{2}}\left(\cos\left(\frac{\theta}{2}\right)e^{-i\alpha\beta/2}\ket{\frac{\alpha}{2}-i\frac{\beta}{2}}\right. \nonumber \\
&\left.+i\sin\left(\frac{\theta}{2}\right)e^{i\phi}\ket{-\frac{\alpha}{2}-i\frac{\beta}{2}}\right)
\end{align}

\begin{align}
\ket{\psi_{+i}}&=\frac{1}{\sqrt{2}}\left(\cos\left(\frac{\theta}{2}\right)\ket{\frac{\alpha}{2}+i\frac{\beta}{2}}\right. \nonumber \\
&\left.-i\sin\left(\frac{\theta}{2}\right)e^{i\phi}e^{-i\alpha\beta/2}\ket{-\frac{\alpha}{2}+i\frac{\beta}{2}}\right)
\end{align}
and $\alpha \beta = \pi /2$. We then apply a CNOD($\gamma$), followed by tomography of the ancilla represented by a partial trace over the EM mode,

\begin{align}
&\mathrm{Tr}_{\mathrm{EM}}\left(D\left(\frac{\gamma}{2}\sigma_{z}\right)\left(\ket{-i}\otimes\ket{\psi_{-i}}-\ket{+i}\otimes\ket{\psi_{+i}}\right)\right. \nonumber \\
&\quad \left.\left(\bra{-i}\otimes\bra{\psi_{-i}}-\bra{+i}\otimes\bra{\psi_{+i}}\right)D^{\dagger}\left(\frac{\gamma}{2}\sigma_{z}\right)\right) = \nonumber \\
&\frac{1}{2}(I + \langle z(\gamma) \rangle\sigma_z + \langle y(\gamma)\rangle\sigma_y + \langle x(\gamma)\rangle\sigma_x)
\end{align}

For $\phi=0$ and $\theta=-\pi/2$ we get $\langle z(\gamma) \rangle = \langle y(\gamma) \rangle= 0$, as expected for the characteristic function of the corresponding cat.
$\langle x(\gamma) \rangle$ is given by,

\begin{align}
&e^{-\frac{1}{2}\left|\gamma\right|^{2}}\left(-\cos\left(\mathrm{Im}\left(\alpha\gamma\right)\right)\sin\left(\mathrm{Im}\left(i\beta\gamma\right)\right)\right. \nonumber \\
&\quad \left.+ e^{-\frac{1}{2}\left|\alpha\right|^{2}}\sinh\left(\mathrm{Re}\left(\alpha\gamma\right)\right)\cos\left(\mathrm{Im}\left(i\beta\gamma\right)\right)\right) \nonumber \\
&+ e^{-\frac{1}{2}\left(\left|\beta\right|^{2}+\left|\gamma\right|^{2}\right)}\cosh\left(\mathrm{Re}\left(i\beta\gamma\right)\right) \nonumber \\
&\quad \times \left(\cos\left(\mathrm{Im}\left(\alpha\gamma\right)\right)+e^{-\frac{1}{2}\left|\alpha\right|^{2}}\cosh\left(\mathrm{Re}\left(\alpha\gamma+i\beta\gamma\right)\right)\right)
\end{align}

As $\alpha$ gets larger, $\langle x(\gamma) \rangle$ gets exponentially closer to the real part of the exact characteristic function,
\begin{equation}
\frac{e^{-\frac{1}{2}\left|\gamma\right|^{2}}}{1+e^{-\frac{1}{2}\left|\alpha\right|^{2}}}\left(\cos\left(\mathrm{Im}\left(\gamma\alpha\right)\right)+e^{-\frac{1}{2}\left|\alpha\right|^{2}}\cosh\left(\mathrm{Re}\left(\gamma\alpha\right)\right)\right)
\end{equation}

The main difference is a small shift in the real axis.

\subsection{Density matrix reconstruction}
We reconstruct the density matrix using a maximum likelihood estimation (MLE) method~\cite{James2001Meaurement}. For measured expectation values $\langle\mathcal{D}(\lambda)\rangle_{\textrm{meas}}$ of a set of $\lambda$'s, the likelihood function to be minimized is given by
\begin{equation}
    \mathcal{L}(\rho_{\mathrm{MLE}}) = \sum_\lambda\frac{\left|\mathrm{Tr}[\mathcal{D}(\lambda)\rho_{\mathrm{MLE}}]- \langle\mathcal{D}(\lambda)\rangle_{\textrm{meas}}\right|}{\delta\langle\mathcal{D}(\lambda)\rangle_{\textrm{meas}}}
\end{equation}
where $\rho_{\textrm{MLE}}$ is the reconstructed density matrix, and $\delta\langle\mathcal{D}(\lambda)\rangle_{\textrm{meas}}$ is the standard error of the measured expectation value.

Once the density matrix $\rho$ is reconstructed, we are able to calculate its fidelity with respect to a target state $\sigma$, according to
\begin{equation}
    \mathcal{F}(\rho,\sigma) = \left(\mathrm{Tr}\left[\sqrt{\sqrt{\rho}\sigma\sqrt{\rho}}\right]\right)^2
\end{equation}

\section{Multi-mode displacements}\label{DIS-FRA}
\subsection{Displaced frame}
It is instructive to analyze the relations of the CNOD gate in the frame of the semi-classical phase space trajectories as defined in Ref. ~\cite{eickbusch2022fast}. Specifically, for a single mode, a transformation by a time dependent unitary $U=D\left(\alpha\left(t\right)\right)=exp\left[\alpha\left(t\right)a^{\dag} -\alpha^*\left(t\right)a \right]$ is applied. As with the ECD gate of Ref.~\cite{eickbusch2022fast}, the CNOD gate includes both the phase space echo $\alpha \rightarrow - \alpha$ and qubit echo $\ket{g}\leftrightarrow \ket{e}$ at time T. Therefore, many of their conclusions for a single mode, such as the significant reduction of the effect of terms proportional to sign$(\alpha)$, can be extended to the CNOD. 

We expand the discussion to a multi-mode setup and consider the effects of higher order terms which include operators of multiple modes, specifically the cross-Kerr term. 
In the two-mode displaced frame, corresponding to a simultaneous application of the CNOD gate on both modes, the cross Kerr term is given by

\begin{align*}
     a^{\dagger}ab^{\dagger}b \rightarrow &  (a^{\dag}+\alpha^{*})(a+\a)(b^{\dag}+\beta^{*})(b+\beta)\\
    = & a^{\dag}ab^{\dag}b+\left|\alpha\right|^{2}b^{\dag}b+\left|\beta\right|^{2}a^{\dag}a\\
    + & \left|\alpha\right|^{2}\beta b^{\dag}+\alpha\left|\beta\right|^{2}a^{\dag}+h.c.
\end{align*}
As with the single-mode higher order terms, we notice that some of the elements in the transformed cross-Kerr term should be greatly reduced by the phase space echoing. The remaining non-negligible terms, $|\alpha|^2 b^{\dagger}b$ and $|\beta|^2 a^{\dagger}a$ correspond in the former (later) to a shift in the frequency of the Bob (Alice) mode which depends only on the displaced frame of Alice (Bob). For the application of CNOD gate to only one of the modes, we notice that this term effects the other mode just by an unconditional rotation of it's phase space.

\subsection{Simultaneous multi-mode CNOD}\label{SimCNODS}

Even with the addition of higher order terms, our effective description of the system with an almost dispersive Hamiltonian is expected to eventually breakdown as the implementations of the CNODs  populate oscillator states with increasing photon numbers ~\cite{PhysRevLett.105.173601,PhysRevLett.117.190503,PhysRevApplied.11.014030}. The expected behaviour is that at large photon numbers higher order transitions, where the transmon ancilla is excited beyond the $\ket{g},\ket{e}$ manifold, are induced. Such higher order nonlinear transitions are expected to limit the Hilbert space accessible to the CNOD and thus reducing the speedup by which it implements conditional displacements.

\begin{figure}[ht!]
    \centering
    \includegraphics[width=1\linewidth]{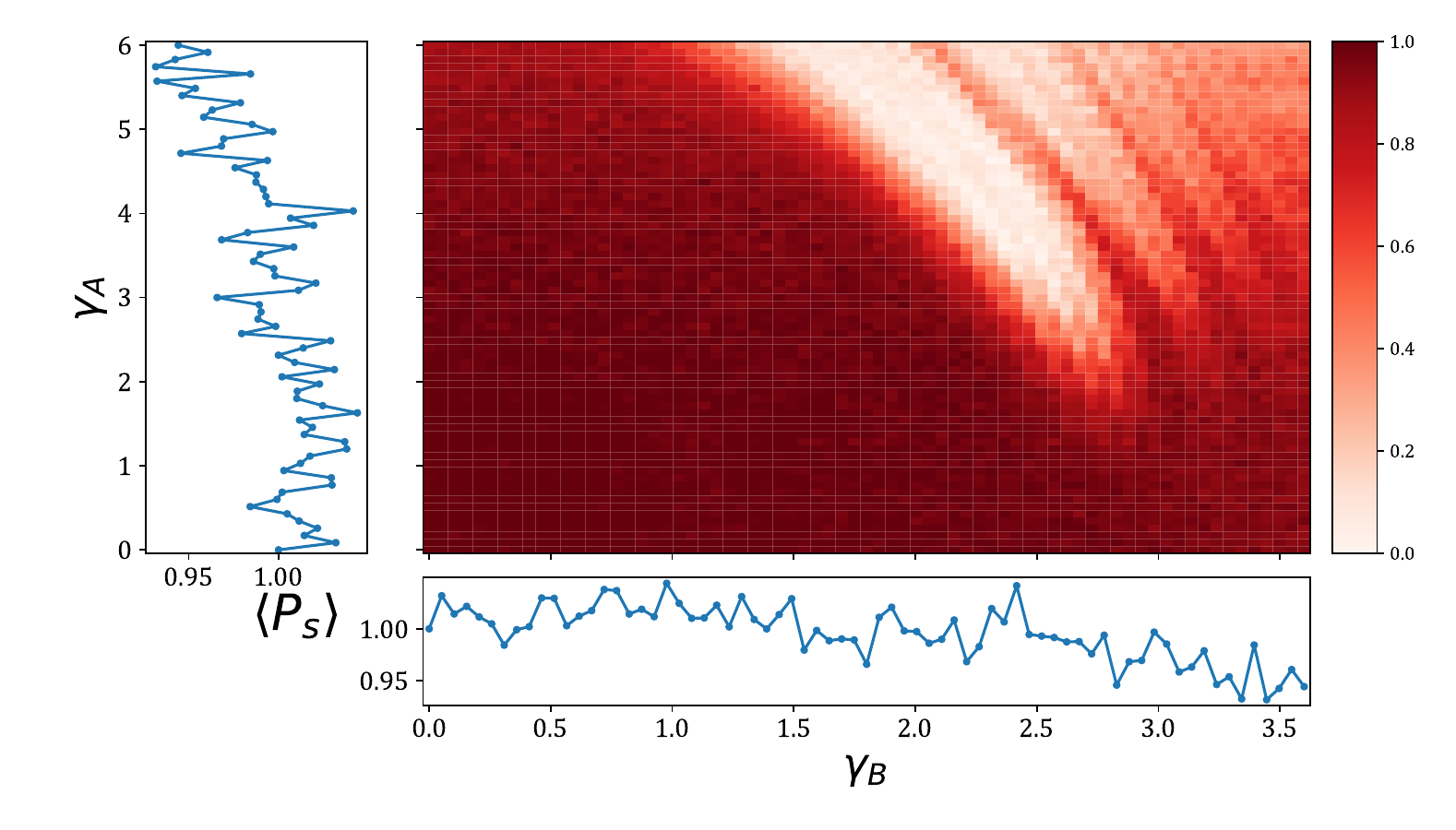}
    \caption{\textbf{Measured normalized purity} $\left\langle{P_s}\right\rangle$ as a function of $\gamma_A$ and $\gamma_B$, the final displacement amplitude and the Alice and Bob mode respectively. The anti-symmetric pulses of which comprise the CNOD are applied simultaneously to the two modes by two pulses with a carrier frequency corresponding to the of each of the modes when the ancilla is excited. The two pulses have the same envelope which is composed of a Gaussian of standard deviation of $\sigma =144$ ns and total length 0f $2\sigma=288$ which is multiplied by $\sin\left(\Delta(t-t_0)\right)$ with $\Delta \approx 10$ Hz and $t_0=144$ ns is half the pulse duration, making anti-symmetric in both time and frequency domain. 
    The results are averaged ~2000 times for each of the tomographic measurements. The results are normalized by the purity at zero displacements to account for the decoherence of the ancilla which is not related to the memory modes.}    \label{fig:cat and back survival}
\end{figure}

To estimate the speed and displacement amplitudes of the various implementations of the CNOD gate at which such additional decoherence is induced, we measure the depolarization of the ancilla during the CNOD gates. We start by preparing the transmon in $(\ket{g}+\ket{e})$ state and apply the conditional 
displacements $\mathrm{CNOD}(\gamma_{A} ,\gamma_{B})$, $\mathrm{CNOD}(-\gamma_{A} ,-\gamma_{B})$, 
with a fast ancilla $\pi$ rotation in between. Finally, a tomographic measurement of the transmon qubit is performed, similar to the measurement of the characteristic function. The normalized purity of the qubit subspace, $\left\langle{P_s}\right\rangle=\left\langle{\sigma_z}\right\rangle^2+\left\langle{\sigma_x}\right\rangle^2+\left\langle{\sigma_y}\right\rangle^2$ is displayed in Fig.~\ref{fig:cat and back survival} as a function of the two-mode displacement amplitudes $\gamma_A$ and $\gamma_B$.

We find that by applying the CNOD to both modes simultaneously the ancilla exhibits a significant depolarization at certain displacement amplitudes. The abruptness of these disappearance lines and the reemergence of the ancilla at larger displacement amplitudes, may indicate that this phenomena is related to resonant transitions to higher levels or residual entanglement with the cavity modes. Indeed, one could use the available sweet spots in the large, simulations displacements for multi-mode state manipulation. However, in this work we chose to implement the large, multi-mode CNODs, such as the first CNOD in the generation of the Bell-cat described in fig. \ref{fig:gates and Setup}b, by a pair of consecutive, single-mode CNODs separated by an ancilla $\pi$ pulse. We postpone to future work an analysis of the physical mechanisms leading to such transitions and if they could be avoided by temporal shaping of the anti-symmetric pulses from which the CNOD is composed.    

On the other hand, we notice that simultaneous application of small conditional displacements does not lead to additional ancila decoherence. Therefore, small, two-mode CNOD displacements, such the second disentangling- CNOD in fig.~\ref{fig:gates and Setup}b required for the generation of the Bell-cat, were applied simultaneously to both modes. 

Furthermore, in the cuts of Fig.~\ref{fig:cat and back survival} dedicated to a single mode application of a CNOD, we find that additional decoherence might occur for the larger displacement. However the effect is very small and the continuous increase in the decoherence indicates that they are most likely related to qubit dephasing induced by photon loss which increase with the size of the overall conditional displacement. A similar mechanism, combined with the absence for the correction of the unconditional phase space rotation discussed in the previous subsection, could explain continuous increase in dephasing observed for very large simultaneous CNODs.

Finally, it is interesting to note that for our current setup and at the range of the single-mode CNOD implementations we used, we did not observe these effects. Something that should be looked into in the future.

\section{Multi-mode Universality}\label{MuMo-UNI}
One way to implement universal control of a single Harmonic oscillator is by producing any Unitary  evolution generated by Hamiltonians polynomial in $X = (1/\sqrt{2})(a^{\dagger}+a)$ and $P = (i/\sqrt{2})(a^{\dagger}-a
)$ \cite{PhysRevLett.82.1784UniCtrThy1,RevModPhys.77.513UniCtlThy2}. By including  Unitaries generated by Hamiltonians of the form $X^jP^k\sigma_i$, \cite{eickbusch2022fast} extends the definition of universal control to include an oscillator and a qubit. In the same work it was shown that the set of operations equivalent to  $\{\mathrm{CNOD}(\vec{\alpha_i}), R(\phi,\theta)\}$ are sufficient to obtain such universal control. 

Extending a Hilbert space comprised of multiple oscillators, universal conrol requires unitaries generated by Hamiltonians compose of arbitrary polynomials of the oscillators ${X_i,P_i}$. However, when universal control can be implemented separately on each oscillator, it is sufficient to include the ability to apply a beam splitter interaction of the form $B_{ij} = P_mX_n-X_mP_n$. 

Working with the generators of the universal control for each single mode,
i.e. $X_{m}^j P_{m}^k \sigma_i$, we expand our set by using the commutators $[X_m^2 \sigma_x,P_n \sigma_y]\propto X_{m}^2 P_n\sigma_z$ for $m\neq n$. To generate the required beam splitter term, which does not include the Pauli operator, we next commute $[X_{m}^2 P_n\sigma_z,X_{m}\sigma_z]\propto X_{m} P_n$. As discussed in the main text, application of the CNOD on a single mode  allows for universal control of each single oscillator, independent of and without effecting the other mode. Thus, with the addition of the single ancilla unconditional  rotation $R_\phi(\theta)$, the combination of the single modes $\mathrm{CNOD}(\gamma_i)$ enables universal control of Hilbert space comprised of multiple oscillators coupled to a single qubit. 

\section{Measurement of non-local operators}\label{NL-OP}
To quantify the entanglement and fidelity  of our state to the Bell-cat, it is sufficient to measure the expectation value of the logical cat-code operator $\mathcal{F}=\frac{1}{4}\left(\left\langle II \right\rangle+\left\langle ZZ \right\rangle+\left\langle XX \right\rangle-\left\langle YY \right\rangle\right)$ which for a general form of the two modes cat-code state $\ket{\psi_G}=a\ket{\alpha_A,\alpha_B}+b\ket{\alpha_A,-\alpha_B}+c\ket{-\alpha_A,\alpha_B}+d\ket{-\alpha_A,-\alpha_B}$ is given by  $\frac{1}{4}\left(|a|^2+|d|^2+ad^*+a^*d\right)$. We proceed to show that by multi-mode CNODs we are able to map the expectation values of the 2-mode operators of the cat qubits onto the single ancilla. 
Starting with the ancilla at the ground state we apply two $\pi/2$ pulses around the Y-axis which are separated by CNOD$\left(i\frac{\pi}{4\alpha_{A}},-i\frac{\pi}{4\alpha_{B}}\right)$ and obtain

\begin{align}
&U_{II+ZZ}\ket{g}\otimes\ket{\psi_G} \approx \ \ket{g}\otimes\left(a\ket{\alpha_{A},\alpha_{B}}+d\ket{-\alpha_{A},-\alpha_{B}}\right) \nonumber \\
&+\ket{-i}\otimes\left(b\ket{\alpha_{A},-\alpha_{B}}\right)+\ket{+i}\otimes\left(c\ket{-\alpha_{A},\alpha_{B}}\right)
\end{align}

where $U_{II+ZZ}=R_{\frac{\pi}{2},\hat{y}}\cdot 
\mathrm{CNOD}\left(i\frac{\pi}{4\alpha_{A}},-i\frac{\pi}{4\alpha_{B}}\right)\cdot R_{\frac{\pi}{2},\hat{y}}$ and we assume $\frac{\pi}{4\alpha_{A}},\frac{\pi}{4\alpha_{A}}<<1$. We conclude by measurement in the Z basis of the ancilla, resulting in the expectation value $\left\langle U_{II+ZZ}^{\dag}\sigma_{z}U_{II+ZZ}\right\rangle =\left|a\right|^{2}+\left|d\right|^{2}$
which is equal the expectation value $\left\langle II \right\rangle+\left\langle ZZ \right\rangle$ of the cat qubits.
Next, a similar procedure where CNOD gate is replaced by CNOD$\left(2\alpha_{A}.2\alpha_{B}\right)$ is use to obtain 
\begin{equation}
\begin{split}
&U_{XX-YY}\ket{\psi_G}= \\
&\	 \ -\frac{1}{\sqrt{2}}\ket -\tens\left(a\ket{2\alpha_{A},2\alpha_{B}}+b\ket{2\alpha_{A},0}+c\ket{0,2\alpha_{B}}\right)\\
& \	 \ -\frac{1}{\sqrt{2}}	\ket +\tens\left(b\ket{0,-2\alpha_{B}}+c\ket{-2\alpha_{A},0}+d\ket{-2\alpha_{A},-2\alpha_{B}}\right)\\
& \	 \ -\frac{1}{2}	\left(\left(a+d\right)\ket g+\left(a-d\right)\ket e\right)\tens\ket{0,0}
\end{split}
\end{equation}
where $U_{XX-YY}=R_{\frac{\pi}{2},\hat{y}}\cdot \mathrm{CNOD}\left(2\alpha_{A}.2\alpha_{B}\right)\cdot R_{\frac{\pi}{2},\hat{y}}$ and we assumed $\alpha_A,\alpha_B$ are large so we may neglect the cross terms. The corresponding measurement of the ancilla results we obtain $\left\langle U_{XX-YY}^{\dag}\sigma_{z}U_{XX-YY}\right\rangle =\frac{a^{*}d+ad^{*}}{2} = \frac{1}{2}\left(\left\langle XX-YY\right\rangle\right)$.

Thus, combining and rescaling results measured following these two sets of operations, we are able to estimate the entanglement of the two cat qubits by
\begin{equation}
    \mathcal{F}= \left\langle U_{II+ZZ}^{\dag}\sigma_{z}U_{II+ZZ}\right\rangle+2\left\langle U_{XX-YY}^{\dag}\sigma_{z}U_{XX-YY}\right\rangle.
\end{equation}
\section{State Fidelty}
\label{sources of infidelty}
The Bell-cats in the table are each compared to the density matrix of the fully mixed logical cat state $\frac{1}{2}\mathcal{D}(\delta)(\ketbra{e^{i\theta_{ph}}\alpha}{e^{i\theta_{ph}}\alpha}+\ketbra{-e^{i\theta_{ph}}\alpha}{e^{i\theta_{ph}}\alpha})\mathcal{D}^{\dagger}(\delta)$ while the product and single-mode states are compared to $\mathcal{N}(\mathcal{D}(\delta)\ket{e^{i\theta_{ph}}\alpha}+e^{i\phi_{rot}}\ket{-e^{i\theta_{ph}}\alpha})$. We did a basic step of optimization, where we allow the cat-states being compared to have small deviations in the parameters $\alpha$, $\delta$, and $\phi$. These could be corrected.

\begin{table}[ht!]

\begin{tabular}{l | l | c c c c c c  }
    Mode
    & State type
    & \textbf{Fidelity}    
    & \boldsymbol{$\alpha$}
    & \boldsymbol{$\delta$} 
    & \boldsymbol{$\theta$}    
    & \boldsymbol{$\phi_{rot}$}  
    \\
\hline
\hline
& Bell-cats &    0.89  &  1.69  &  -0.02-0.01i  &  -  & 0.00 \\
Alice & product-cats  &  0.83 & 1.67 & 0.01+0.06i & 0.21 & 0.00 \\
& single-cat &  0.87 & 1.68 & -0.04+0.05i & 0.336 & 0.00 \\
\hline
& Bell-cats & 0.92  & 1.75   & -0.16 +0.05i & - & 0.00 \\
Bob & product-cats & 0.77 & 1.71  &  -0.1+0.05i & -0.06 & 0.25 \\
& single-cat  &  0.86 & 1.75 & -0.1+0.02i & 0.02  & 0.12 \\
\hline

\end{tabular}

\caption{\textbf{Fidelity of prepared states}}

\label{table:fidelty}
\end{table}

A full simulation could yield a transfer matrix for the process, and separate out the contribution of the different errors. However, separating and giving an exact quantitative analysis and comparing an error budget is challenging at this point. Ideally one would simulate the full system including measurements to extract the contributions, however the large Hilbert space makes it prohibitive, and we are looking into numerical methods to do this. We can get a qualitative understanding of the error budget and separate the contributions of some sources of infidelity from the data and from simple simulations assuming perfect unitaries. There are 4 major such sources, qubit decoherence; photon loss; finite size $\chi$ effects, such as self-Kerr; and imperfections of the scheme, mostly the imperfect disentanglement from the qubit. 

We perform a gate-based simulation by applying the calculated unitaries, therefore neglecting the effects of SPAM errors, qubit decoherence, photon loss, and finite size $\chi$. Without all these errors the primary source of infidelity is imperfect disentanglement, which improves with cat size. Figure \ref{fig:disenterror} shows simulation results for creating a single cat state as function of cat size, which sets an upper bound on the created cat states. For the cat size in our experiment the simulation shows a fidelity of 0.95. At a cat size of 4 photons, the upper bound on fidelity exceeds 0.99. Employing additional operations that better approximate the state could further enhance fidelity.

\begin{figure}[htp!]
    \centering
    \includegraphics[width=1\linewidth]{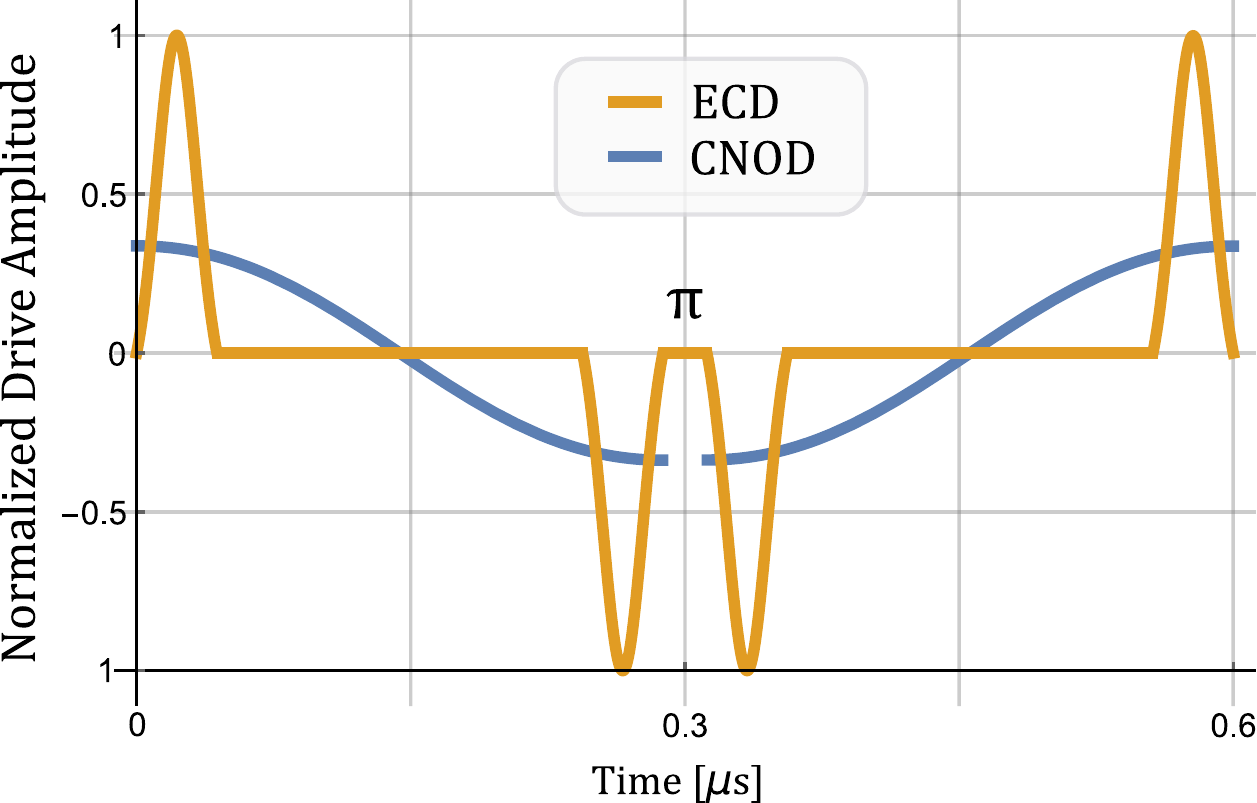}
    \caption{\textbf{A comparison between drive amplitudes of CNOD and ECD.} 
    } 
    \label{fig:CNOD-ECD}
\end{figure}

We can gain qualitative insight into the dominant sources of infidelity by comparing the fidelities obtained for single-mode states and product-cat states in the two modes. This comparison reveals that the infidelity is primarily influenced by single-photon loss and transmon decoherence. To generate the product-cat states, we first create a cat state in the Bob mode, followed by generating a cat state in the Alice mode. Then, we perform characteristic tomography, specifically measuring the desired mode. The preparation of the Alice mode is affected by transmon decoherence during the preparation of the Bob mode, resulting in increased initialization errors. This likely contributes to the reduction in fidelity from 0.87 to 0.83 when comparing the product-cat state to the single-cat state. Since the Bob mode is prepared first, single-photon loss in the Bob mode during the preparation of the Alice cat state leads to a decrease in measured fidelity. This is due to the additional time waited until tomography is performed when comparing the product-cat state to the single-cat state. Furthermore, transmon decoherence also contributes to a further decrease in fidelity. This analysis is supported by the more significant decrease in fidelity observed between the single- and product-cat states of the Bob mode compared to the Alice mode.

\section{Comparing the CNOD and ECD methods}
A qualitative comparison between our CNOD method and the recently demonstrated ECD method~\cite{eickbusch2022fast} would be instructive, as both approaches employ a large photon number to reduce the gate duration and essentially yield the same unitary transformation (with a minor variation in the axis of the intermediate $\pi$ pulse). Since the gate speed in our experiment is constrained by the maximum driving amplitude, we compare the peak power requirements of the two methods. As an illustrative example, in Fig.~\ref{fig:CNOD-ECD}, we use the first pulse applied to the Bob mode to create the cat state and plot the ECD gate with the same duration needed to achieve the same target state in the setup used in this work, employing pulse parameters similar to those used in Ref.~\cite{eickbusch2022fast} (displacement with a Gaussian profile having $\sigma$=11ns and pulse duration of 44ns). We find a difference of 9.44 dB in the required peak power. 

In ref~\cite{eickbusch2022fast}, the scaling of the displacement amplitude given in the manuscript is $\propto A \chi \tau$, where A is the typical amplitude of the delta functions of the ECD. The ECD is analysed in terms of the delta functions composing it, and $\tau$ is the \emph{wait time} between the delta functions. In the CNOD the scaling of the displacement amplitude is $\propto A \chi \tau^2$. This would suggest favorable scaling of the CNOD vs the ECD. However, in practice, the delta functions are Gaussians. If we consider their duration to also scale, then the ECD displacements amplitude will also scale in time with $\tau^2$. This suggests that the precise pulse shape makes a difference for both methods, and the ECD analysis using delta functions may suffer from the accuracy of the approximation. In this sense we believe that the CNOD concept, which only requires the anti-symmetry, is simpler to work with and apply in practice.

When optimizing, an additional particularly intriguing aspect to consider is the limitation on gate speed imposed by higher-order transitions and the breakdown of the dispersive approximation at very high photon numbers, such as cavity-qubit ionization, sometimes referred to as bright stating \cite{lescanne_detecting_2019,ShillitoPhysRevApplied.18.034031}. As both approaches employ intermediate states with a significant number of photons to achieve shorter gate durations, a maximum photon number could potentially restrict the gate speed. While an estimate of the speed limit can be derived from the critical photon number at which the dispersive approximation breaks down due to higher-order effects~\cite{blais2021circuit}, it should be regarded as a guideline rather than an exact threshold. 
In fact, we did not observe any evidence of such transitions in the single-mode CNODs (see Fig.~\ref{fig:cat and back survival}), this is contrast to the observations reported in ~\cite{eickbusch2022fast}. The reason for this discrepancy is currently unknown and could potentially be attributed to differences in the experimental setups, including system parameters. Another possible explanation may be found in the description of this phenomenon in terms of Landau-Zener transitions \cite{ShillitoPhysRevApplied.18.034031} Unlike the ECD gate, the continuous driving during the CNOD gate could aid in suppressing such transitions by rendering them primarily diabatic. However, a definitive conclusion would necessitate a more rigorous comparison of the two methods. 

There exist additional distinctions between the two methods, such as the ECD gate not being perfectly anti-symmetric, leading to more subtle differences. A comprehensive comparison of the two methods, such as studying the limitations imposed by large photon numbers and potentially offering a unified framework for the two methods, is left for future work. This analysis will go beyond peak power comparisons and provide a more thorough understanding of the similarities and differences between the ECD and CNOD methods.

\bibliography{references}

\end{document}